\documentclass{article}

\usepackage{arxiv}

\usepackage[utf8]{inputenc}
\usepackage[T1]{fontenc}
\usepackage{hyperref}
\usepackage{url}
\usepackage{booktabs}
\usepackage{amsfonts}
\usepackage{microtype}
\usepackage{graphicx}
\usepackage{natbib}
\usepackage{doi}

\usepackage{amsmath}
\usepackage{amsthm}
\usepackage{mathtools}
\usepackage{dsfont}
\usepackage{array}
\usepackage{subcaption}
\usepackage{float}
\usepackage{longtable}
\usepackage{algorithm}
\usepackage{algpseudocode}
\usepackage{appendix}
\usepackage{adjustbox}
\usepackage{xcolor}
\usepackage{hhline}
\usepackage{rotating}
\usepackage{wrapfig}
\usepackage{sidecap}

\allowdisplaybreaks[2]

\theoremstyle{definition}
\newtheorem{definition}{Definition}[section]

\theoremstyle{remark}

\title{Beyond the Fixed Price: Valuation and Risk of Non-Standard Renewable PPAs}

\author{
Nicola Bartolini\thanks{Corresponding author: nicola.bartolini11@unibo.it}\\
Department of Statistics\\
University of Bologna\\
Bologna, Italy\\
\texttt{nicola.bartolini11@unibo.it}
\And
Silvia Romagnoli\\
Department of Statistics\\
University of Bologna\\
Bologna, Italy \\
\texttt{silvia.romagnoli@unibo.it}
\And
Amia Santini\\
Department of Statistics\\
University of Bologna\\
Bologna, Italy\\
\texttt{amia.santini2@unibo.it}
}

\hypersetup{
pdftitle={Beyond the Fixed Price: Valuation and Risk of Non-Standard Renewable PPAs},
pdfauthor={Nicola Bartolini, Silvia Romagnoli, Amia Santini},
pdfkeywords={Power Purchase Agreements, Renewable Energy, Risk Management, Monte Carlo Simulation, Electricity Markets}
}

\begin{document}

\maketitle

\begin{abstract}
Renewable Power Purchase Agreements have become increasingly important instruments for supporting the energy transition, as they offer revenue stability to renewable energy producers and price certainty to electricity consumers. This paper develops a financial framework for the valuation and risk assessment of fixed-price renewable PPAs. We formalize the payoff structures of the main PPA designs adopted in practice for wind and photovoltaic generation and derive fair contract prices based on financial valuation principles. We further propose a market risk-assessment methodology based on Monte Carlo simulation and introduce a parsimonious continuous-time model for solar irradiance suitable for financial applications. An empirical analysis of the Italian electricity market shows that fair prices and risk profiles vary substantially across technologies and contractual structures, highlighting the trade-off between downside protection and participation in favorable market outcomes. This framework provides practical tools for the pricing and risk evaluation of renewable PPAs.
\end{abstract}


\section{Introduction}
Renewable Power Purchase Agreements (PPAs) are contracts between a producer of electricity from renewable energy sources and an offtaker. They formalize an agreement in which the offtaker commits to the periodic purchase of a fixed proportion (or the totality) of the power output of the producer for a pre-determined price. This price could be fixed or floating, varying based on the market price of electricity, and could be bounded - above or below - depending on the specific requirements of the two parties (\cite{Hundt202135}). Such contracts are bound to play a key role in the energy transition, as they allow for the two parties to reduce the economic uncertainty related to their business.
\\
The primary benefit for the producer is financial in nature. Renewable energy projects require very large upfront investments, but have comparatively lower operating costs than other energy sources. A PPA guarantees a fixed or bounded price for the electricity produced over the duration of the contract (typically 10-20 years). This predictable cash flow can act as the collateral that allows developers to secure financing of an otherwise unbankable project. It also protects producers against the risk of drops in future wholesale power prices, ensuring that they can cover their debt obligations and operational costs (\cite{POMBOROMERO2024107861}, \cite{baringa2022commercial}). 
For the offtaker, the benefits involve economic stability and strategic positioning. PPAs allow large energy consumers to lock in a fixed electricity rate for years, effectively hedging against the kind of price spikes seen during the 1970s oil crisis or the recent European energy crisis. Signing a renewable PPA is also a key tool for corporations to verifiably reduce their carbon footprint (Scope 2 emissions) and meet public ESG commitments for renewable energy usage.
\\
However, the over-the-counter nature of PPAs and their multi-year maturity exacerbate the risks associated with them: credit risk, volumetric risk, and price risk. Recognizing the barrier that these pose to PPA adoption, particularly for smaller corporate offtakers, the European Investment Bank (EIB) has designed a pilot guarantee scheme, which was announced in early 2025. This project has an indicative budget of EUR 500 million\footnote{European Investment Bank, ``PAN-EU POWER PURCHASE AGREEMENT GUARANTEE", Project No. 20250202, approved 19 June 2025. \url{https://www.eib.org/en/projects/all/20250202}} and works by shifting to the EIB part of the credit risk carried by a third-party financial intermediary issuing a guarantee to the renewable energy developer on behalf of the corporate buyer.
\\
In this work, we aim to support the EIB guarantee scheme by providing a financial framework for the valuation and risk assessment of such instruments. We do so via four main contributions: first, we formalize, in financial mathematics terms, the payoff structure of the fixed-price renewable PPA types that have been identified in \cite{wbcsd_pricing_2021_alt}, for wind and for photovoltaic (PV) power production. Second, for each type, we provide a pricing approach relying on the financial concept of fairness rather than on the Levelized Cost of Electricity, thus extending the framework of \cite{PENA2026125168} to multiple PPA structures. In doing so, we also allow for settlement and delivery periods of the contracts not to coincide, in order to be able to represent most publicly available contracts (\cite{wbcsd_pricing_2021_alt}). With our contribution, we aim to provide the regulator a tool to detect and quantify deviations from market fairness, so that proper action can be taken to understand the reasons behind them or to correct potential inefficiencies. 
\\
Third, partially addressing the research gap highlighted by \cite{zhao2026hybrid}, we illustrate a risk-assessment methodology conceptually aligned with current financial disclosure standards and able to balance accuracy with the complexity of the PPA that is caused by the large number of time-steps in the contract lifetime and by the interaction between the underlying risk factors. In fact, the complex nature of the instruments does not permit to reach closed-form pricing formulas but requires the reliance on time-consuming and computationally-demanding Monte-Carlo methods. 
\\
Lastly, we introduce a continuous-time model for the Global Horizontal Irradiance, the key variable behind PV power production, selected in line with the features of the data and parsimonious enough to be compatible with the financial setting of our work. 
\\
The remainder of this paper is structured as follows: Section \ref{sec:literature_review} contains the literature review, clearly locating our contributions in the existing landscape, Section \ref{sec:general_framework} presents the PPA pricing framework and the continuous-time models for the underlying stochastic quantities, Section \ref{sec:model_selection} contains information about the data sources and about the calibrated model parameters, Section \ref{sec:empirical_analysis} illustrates an empirical application of the pricing and risk-assessment methodologies to the Italian market. Finally, Section \ref{sec:conclusions} concludes.

\section{Literature review}
\label{sec:literature_review}
The current literature on PPAs covers a number of different topics. 
The first direction which is relevant for the present work concerns the valuation of PPAs. \cite{MENDICINO2019113577}, propose the use of the Levelized Cost of Energy formula (LCOE), an engineering-based approach to evaluate PPAs, with the aim of identifying the optimal contract duration. They find it to be between 7 and 10 years. The first instance of a financial approach to PPA pricing is developed by \cite{PENA2026125168}. In the present work, we build on the authors' reliance on the concept of fairness and extend it beyond the case of the fixed-price PPA, covering more complex contractual structures.
\\
Another common thread running through recent PPA literature concerns financial feasibility, which is found to depend on contract structure, financing conditions, and market dynamics. \cite{cuervo2021photovoltaic} study the financial feasibility of a PPA that uses photovoltaic energy and find that it is possible to develop a business model based on photovoltaic energy in Colombia only when the tax benefits of Law 1715 are considered. From the point of view of the flexibility of the investment, the value of the option to defer is greater than the NPV obtained from the discounted cash flow model, indicating the convenience of delaying the investment in renewable energy.
\cite{simoes2025power} show that baseload contracts consistently
outperform other structures during periods of price
cannibalization, offering the best equilibrium between risk mitigation and performance. \cite{kandpal2024power} develop a framework for continuous 24/7 PPA contracts with an hourly renewable energy delivery timeline, which proves effectively manage the hour-to-hour power delivery commitments 
\\
As for the quantification of PPA risk, \cite{hundt2021power} show that the PPA price is strongly influenced by a firm's financial leverage and vice versa, because the debt sizing strongly depends on the offtaker’s creditworthiness. Their results show that offtakers with a lower creditworthiness have limited access to PPA markets because of higher PPA prices that are necessary to compensate the owner of the RE asset for the higher default risk. Furthermore, \cite{zhao2026hybrid} review the emerging market trends and current practices in PPA procurement and show that while the literature provides extensive qualitative discussions of 24/7 PPA design, quantitative modeling frameworks remain limited. In particular, they say that an integrated framework that jointly captures PPA pricing, contract quantities, operational flexibility, and risk management is still lacking, although there are many works describe PPA structures and characteristics. Our work aims to address this gap.
A study of the risks associated with solar PPA contracts has been made by \cite{simoes2025efficient}, but with a different focus than what we do in this work. \cite{simoes2025efficient} evaluate the impact of the different combinations of electricity profile (Pay-as-Produced, Fixed Hourly Profile, Monthly Baseload, and Annual Baseload) and price structure (Fixed Price and Variable Price). They find that among solar PPAs, contracts with a monthly baseload profile and a variable price structure achieve the highest overall performance.

\section{General Framework}
\label{sec:general_framework}
The framework in this study involves a producer of renewable energy, either from wind or photovoltaic power, and an offtaker. The two parties sign a PPA at a time $t$ for the sale and purchase of a fixed percentage (or the entirety) of the amount of output $Q(T_i)$ generated by the producer during each \textit{delivery period} $T_i$, $i=1,...,n$, with $T=T_n$ being the final maturity of the agreement. The price per MW/h is fixed at a level $K$, with potentially additional provisions on its value, and is exchanged on every \textit{settlement} date. The length of the delivery period does not necessarily correspond to that of the settlement period, with the latter usually being a multiple of the former (\cite{wbcsd_pricing_2021_alt}). 

The risks connected to the contract involve the produced quantity during every period $T_i$, $Q(T_i)$, which is stochastic and driven by either wind speed or by Glbal Horizontal Irradiance (GHI), as well as the market value of the electricity being delivered, which we denote by $S(T_i)$.
PPAs can differ in their structure, delivery profile, pricing indexation, settlement frequency, tenor, and credit support mechanisms. In the remaining part of this section, we formalize, in financial-mathematical terms, the payoff structure of the fixed-price contracts identified in the technical report of \cite{wbcsd_pricing_2021_alt} and, for each type, we provide a pricing approach relying on the financial concept of fairness.

\subsection{Contract definition and risk-neutral pricing}
\label{sec:contract_definitions_and_risk_neutral_pricing}
In this section, we tackle the first two objectives of this paper: we propose formal definitions of the different existing fixed-price renewable PPA types that have been identified in \cite{wbcsd_pricing_2021_alt} and \cite{MITTLER2025115293}. Then, for each type, we provide a pricing approach that relies on the financial concept of fairness rather than on the Levelized Cost of Electricity, thus extending the framework of \cite{PENA2026125168} to multiple PPA structures. We choose the financial mathematics approach in light of the results of \cite{PENA2026125168}, who perform empirical tests on utility-scale solar and wind PPAs and show that the financial-based pricing approach outperforms traditional LCOE approaches, with the additional advantage of transparency and replicability. We further allow for settlement and delivery periods of the contracts not to coincide, in order to be able to cover a wider range of existing contracts.

Due to the current lack of liquid financial derivatives contracts written either on $Q(T)$ or on the risk factors associated with it, we assume, as \cite{PENA2026125168}, that the risk-neutral pricing measure $\mathbb{Q} = \mathbb{P}$ for the renewable energy output.

\begin{definition}
\label{def:payoff_ppa_plain_vanilla}
\textbf{Fixed-price PPA payoff}\newline
The payoff to the offtaker of a fixed-price PPA signed at time $t$ with maturity $T=T_{m_n}$, delivery periods $T_j$, $j=1, ... , m_i$, $i=1,...,n$, settlement periods $T_{m_i}$, $i=1,...,n$, and fixed price $K$ is
\begin{equation}
        \label{eq:plain_vanilla_ppa_payoff}
        \Pi(t, T) = \sum_{i=1}^{n} \sum_{j=1}^{m_i} Q(T_j)\big(S(T_j) - K\big)
    \end{equation}
where $Q(T_j)$ is the quantity of electricity produced at $T_j$ and $S(T_j)$ is the price of electricity at $T_j$.
\end{definition}

\begin{definition}
\label{def:fair_price_ppa_plain_vanilla}
\textbf{Fixed-price PPA risk-neutral price}\newline
The risk-neutral price of a PPA is the constant payment $K \in \mathbb{R}^+$ that makes the discounted risk-neutral conditional expectation of the PPA payoff equal to zero at contract inception $t$. For a fixed-price PPA, this is
\begin{equation}
    \label{eq:ppa_fair_price_generic}
    K = \frac{\sum_{i=1}^n \mathbb{E}_t^\mathbb{Q}[\sum_{j=1}^{m_i} Q(T_j)S(T_j) D(t,T_{m_i})]}{\sum_{i=1}^n \mathbb{E}_t^\mathbb{Q}[\sum_{j=1}^{m_i} Q(T_j)D(t,T_{m_i})]}.
\end{equation}
where $D(t, T_{m_i})$ is the discount factor from time $T_{m_i}$ to time $t$.
\end{definition}

\begin{definition}
\label{def:stepped_ppa_payoff}
\textbf{Stepped PPA payoff}\newline
The payoff to the offtaker of a stepped PPA, i.e. a fixed-price PPA with escalation, signed at time $t$ with maturity $T=T_{m_n}$, delivery periods $T_j$, $j=1, ... , m_i$, $i=1,...,n$, settlement periods $T_{m_i}$, $i=1,...,n$, and fixed base price $K$ is
\begin{equation}
        \label{eq:stepped_ppa_payoff}
        \Pi(t, T) = \sum_{i=1}^{n} \sum_{j=1}^{m_i} Q(T_j)\Big(S(T_j) - K\times\big(1+f(S,T_j)\big)\Big),
    \end{equation}
where $f(S,T_j)$ is a step function that increases or decreases $K$, depending on time or on different drivers, such as the price of electricity.
\end{definition}

\begin{definition}
\label{def:stepped_ppa_contract}
\textbf{Stepped PPA risk-neutral price}\newline
The risk-neutral price of a stepped PPA is the constant payment $K \in \mathbb{R}^+$ that makes the discounted risk-neutral conditional expectation of the stepped PPA payoff equal to zero at contract inception $t$, i.e.
\begin{equation}\label{eq:stepped_ppa_price}
    K = \frac{\mathbb{E}^\mathbb{Q}_t[\sum_{i=1}^{n} \sum_{j=1}^{m_i} Q(T_j)S(T_j) D(t,T_{m_i})]}
    {\mathbb{E}^\mathbb{Q}_t[\sum_{i=1}^{n} \sum_{j=1}^{m_i} Q(T_j)\times(1+f(S,T_j))D(t,T_{m_i})]},
\end{equation}
where $D(t, T_{m_i})$ is the discount factor from time $T_{m_i}$ to time $t$.
\end{definition}

\begin{definition}
\label{def:reverse_collar_PPA_payoff}
\textbf{Reverse collar PPA payoff}\newline
The payoff to the offtaker of a reverse collar PPA, signed at time $t$ with maturity $T=T_{m_n}$, delivery periods $T_j$, $j=1, ... , m_i$, $i=1,...,n$, settlement periods $T_{m_i}$, $i=1,...,n$, and fixed price $K$ is
\begin{align}
    \label{eq:ppa_payoff_reverse_collar}
    \nonumber
    \Pi(t, T) = \sum_{i=1}^{n}\sum_{j=1}^{m_i} Q(T_j) & \Big(\big(S(T_j) - K_{max} \big)^+ -\big(K_{min} - S(T_j)  \big)^+ \\
    & + \big(S(T_j) - K \big)\mathds{1}_{S(T_j)\in (K_{min}, K_{max})}\Big),
\end{align}
where $K_{min}< K < K_{max}$ and $K_{min}, K, K_{max}\in \mathbb{R}^+$. The offtaker provides the power producer with a floor price $K_{min}$, guaranteeing a minimum price, while the producer provides the offtaker with a cap price $K_{max}$, safeguarding them from price spikes.
\end{definition}

\begin{definition}
\label{def:reverse_collar_PPA}
\textbf{Reverse collar PPA risk-neutral price}\newline
The risk-neutral price of a reverse collar PPA is the constant payment $K \in \mathbb{R}^+$ that makes the discounted risk-neutral conditional expectation of the reverse collar PPA payoff equal to zero at contract inception $t$, given the contractually agreed-upon bounds $K_{min}$ and $K_{max}$, i.e.
\begin{align}
\label{eq:fair_strike_collar_ppa}
K&=
\Big(
\displaystyle
\sum_{i=1}^{n}\sum_{j=1}^{m_i}
\mathbb{E}^\mathbb{Q}_t\big[
Q(T_j)
\big((S(T_j)-K_{max})^+ -(K_{min}-S(T_j))^+ \notag \\
&\quad + S(T_j)\mathds{1}_{S(T_j)\in (K_{min}, K_{max})} \big)D(t, T_{m_i}) \big] \Big) 
\notag \\
& \quad \Big/   \Big( \sum_{i=1}^{n}\sum_{j=1}^{m_i}
\mathbb{E}^\mathbb{Q}_t\big[
Q(T_j)\mathds{1}_{S(T_j)\in (K_{min}, K_{max})} D(t, T_{m_i}) \big] \Big),
\end{align}
where $D(t, T_{m_i})$ is the discount factor from time $T_{m_i}$ to time $t$.
\end{definition}

In this work, we assume interest rates to be deterministic, so that the focus can be shifted towards addressing the other two sources of risk, which are specific to this instrument: volumetric risk and price risk.

\subsection{The risk factor models}
\label{sec:model}
In this section, we first formally define the functional forms linking each climate variable to the corresponding power production function $Q(T_j)$. Next, we illustrate the continuous-time stochastic processes used to model the underlying risk factors of the PPA.
\\
The renewable contracts in this work are based on wind and on PV power production. In the former case, the amount of electricity produced in any period $T_j$, $Q(T_j,W)$, is a function of the average wind speed for that period, $W(T_j)$. In the latter case, the amount of electricity produced in any period $T_j$, $Q(T_j, GHI)$, is a function of $GHI(T_j)$, the average Global Horizonatal Irradiance of the period. 

\begin{definition}{\textbf{Wind power production.}}
Following \cite{risks6020056}, wind power production $Q(T_j,W)$ is represented as the following function of the average wind speed of period $T_j$, $W(T_j)$
\begin{equation*}
Q(T_j,W) = h W^3(T_j) \mathds{1}_{m \le W(T_j) \le M},
\end{equation*}
where $h$ is the heat rate, $m$ is the minimum wind speed for power production and $M$ is the maximum one.
\end{definition}
For simplicity, without loss of generality, we consider $h=1$ for the purpose of this work, offering unitary output values.
\\
As for PV power, we adopt a linear specification based on the findings of \cite{ASTE2013503} and \cite{LARSON201611}, who show an approximately linear relationship between PV production and GHI.
\begin{definition}{\textbf{PV power production.}}
PV power production is a linear function of $GHI$, i.e. 
$$Q(T_j, GHI)= \alpha GHI(T_j),$$ 
where $\alpha \in \mathbb{R}^+$ aggregates installed capacity, average performance ratio, conversion efficiency, and the mean tilt-related transposition factor. This retains the first-order sensitivity of PV output to irradiance, while meeting the parsimony needs of financial modeling.
\end{definition}

As in the case of wind contracts we consider a unitary scaling constant, in this case $\alpha=1$, without loss of generality.
\\
The stochastic processes which drive the uncertainty in the PPA contracts in this work are three: the spot price of electricity, average daily wind speed, and average daily $GHI$. For each one, a continuous-time stochastic process is selected, on the basis of the features of the corresponding time series. The analyses and selection procedures are presented in Section \ref{sec:model_selection}, while in this section we only present the final models for each quantity.
\\
The spot price of electricity is modeled with a deterministic seasonal component, $\Lambda^S_t$, and an Ornstein-Uhlenbeck process $X_t$ having stochastic volatility $\nu_t$, following established approach of \cite{geman2005commodities}. The parameters, as well as the latent variance process, are calibrated following the procedure illustrated in \cite{alfonsi2024stochastic}.

\begin{align}
    \label{eq:electricity_price}
    & S_t = \Lambda^S_t + X_t 
    \\
    \label{eq:seasonal_part_electricity_price}
    & \Lambda^S_t = \sum_{k=1}^N a^S_k \sin\bigg(\frac{2\pi t}{365} \bigg) + \sum_{k=1}^M b^S_k \cos\bigg(\frac{2\pi t}{365} \bigg) 
    \\
    \label{eq:non_seasonal_part_electricity_price}
    & dX_t = \alpha^X(\bar{X} - X_t)dt + \sqrt{\nu^X_t} dB_t^{X} 
    \\
    \label{eq:volatility_electricity_price}
    & d\nu^X_t = \beta(\bar{\nu}^X - \nu^X_t)dt + \eta^X\sqrt{\nu^X_t} dB_t^{\nu^X} 
    \\
    & \mathbb{E}[dB_t^{X} dB_t^{\nu^X} ] = \rho_{x,\nu} dt
\end{align}

For the wind speed we use an additive model with a seasonal component $\Lambda^W_t$ and a Cox-Ingersoll-Ross process $Y_t$, following \cite{Bensoussan20161355}, ensuring only positive wind speed values. The wind speed process and the spot price process are assumed to be correlated via Eq. \eqref{eq:correlation_wind}.

\begin{align}
    \label{eq:wind_speed}
    & W_t = \Lambda^W_t + Y_t 
    \\
    \label{eq:seasonal_part_wind}
    & \Lambda^W_t = \sum_{k=1}^N a^W_k \sin\bigg(\frac{2\pi t}{365} \bigg) + \sum_{k=1}^M b^W_k \cos\bigg(\frac{2\pi t}{365} \bigg) 
    \\
    \label{eq:non_seasonal_part_wind}
    & dY_t = \kappa(\bar{Y} - Y_t)dt + \sigma_y\sqrt{Y_t} dB_t^{W}
    \\
    \label{eq:correlation_wind}
    & \mathbb{E}[dB_t^{W}dB_t^{X}] = \rho_{x,w} dt
\end{align}

Finally, we model the solar irradiance $GHI_t$ in Eq. \eqref{eq:solar_irradiance} as the product of a deterministic function of space and time, $f(\theta,t)$, representing an upper bound of the clear-sky values, and of a stochastic re-scaling component, $Z_t \in (0,1)$. The stochastic component is then modeled via its logit transform in Eq. \eqref{eq:logit_part_solar}, as the sum of a continuous-time stochastic process, $G_t$, and of a seasonal component, $\Lambda^G_t$. The stochastic process $G_t$ in Eq. \eqref{eq:ou_part_solar} is mean-reverting with stochastic volatility $\nu^G_t$. The two Brownian motions $B^G_t$ and $B^{\nu^G}_t$ are correlated, and $B^G_t$ is also correlated with the Brownian motion driving the electricity spot price, $B^X_t$.

\begin{align}
    \label{eq:solar_irradiance}
    & GHI_t = f(\theta,t) \times Z_t
    \\
    \label{eq:logit_part_solar}
    & Z_t = \frac{1}{1+e^{- (\Lambda^G_t +  G_t)}}
    \\
    \label{eq:ou_part_solar}
    & dG_t = \alpha^G(\bar{G} - G_t)dt + \sqrt{\nu^G_t} dB_t^{G} 
    \\
    \label{eq:volatility_solar}
    & d\nu^G_t = \beta^G(\bar{\nu}^G - \nu^G_t)dt + \eta^G\sqrt{\nu^G_t} dB_t^{\nu^G} 
    \\
    & \mathbb{E}[dB_t^{G} dB_t^{\nu^G} ] = \rho_{g,\nu^{G}} dt
    \\
    & \mathbb{E}[dB_t^{G} dB_t^{X} ] = \rho_{g,x} dt
    \\
    \label{eq:seasonal_part_solar}
    & \Lambda^{G}_t = \sum_{k=1}^N a^G_k \sin\bigg(\frac{2\pi t}{365} \bigg) + \sum_{k=1}^M b^G_k \cos\bigg(\frac{2\pi t}{365} \bigg) 
\end{align}

The function $f(\theta,t)$ is a rescaled version of the deterministic model of clear-sky GHI in \cite{haurwitz1945insolation} and \cite{haurwitz1946insolation} and is described in Section \ref{sec:solar_irradiance_data}. This model was found by \cite{reno2012global} to have a comparable performance to alternative stochastic models and to provide the best performance among the deterministic models present in the literature. It only depends on sun position $\theta_t$, which is a function of time and of the latitude, longitude, and altitude of the location in which the irradiance is measured. We perform a rescaling of the function to ensure that its values are never exceeded, so that it can serve as an upper bound for $GHI_t$. More details on the procedure are presented in Section \ref{sec:solar_irradiance_data}.
\\
\subsection{Contract risk-assessment}\label{sec:risk_assessment}
The third aim of this work is to propose a risk-assessment methodology for the value of PPAs which is conceptually aligned with current prudential financial disclosure standards and that balances accuracy with the complex nature of such contracts. Such complexity is caused by the large number of time-steps in the contract lifetime and by the interaction between the underlying risk factors.
\\
As already highlighted in \cite{zhao2026hybrid}, a framework for the quantitative evaluation of the contracts is still missing in the current literature. In this section we aim to address this gap and introduce how to evaluate the riskiness of PPAs. When taking about the riskiness in their value, it is not fully possible to replicate the assessment methodology of market risk in standard financial markets, and so in terms of daily Mark-to-Market VaR (Value at Risk) and ES (Expected Shortfall). These two metrics, reflecting the potential negative change in the market value of an instrument, find their most straightforward application with standardized contracts exchanged on dematerialized markets. On the contrary, PPAs are OTC contracts, highly customized to the needs of the two involved parties. There is no daily portfolio re-balancing or allocation, as the contract aims to regulate transactions, and often the physical delivery of electricity, for a medium to long time horizon.
\\
We therefore shift the focus to the expected contract payoff at maturity, as it can be evaluated without requiring market trades of an identical contract and as it reflects the projected impact of the agreement on the balance sheet of the signing parties. Therefore, instead of a daily Mark-to-Market VaR and ES, in the absence of a market, we propose the valuation of Terminal VaR and ES, estimating them as the 5th percentile and average tail value of the simulated terminal profit and loss distribution. The complex nature of PPAs does not permit to reach closed-form pricing formulas: in turn, the risk-assessment procedure needs to rely on Monte-Carlo methods. A summary of the procedure is presented in Appendix \ref{app:algorithm_var_es} in algorithm form, for all PPA types under consideration and for both technologies (wind and PV).
An illustration of this analysis is presented in Section \ref{sec:payoff_analysis}.

\section{Model selection and fit}\label{sec:model_selection}

\subsection{Data}
\label{sec:data_section}
The spot electricity price data used in this work corresponds to the time series of the Italian unique national price (PUN), which can be downloaded freely from the GME (Gestore del Mercato Elettrico) at \url{https://gme.mercatoelettrico.org/it-it/Home/Esiti/Elettricita/MGP/Esiti/PUN}. As for the average daily wind speed, the data has been downloaded from \url{https://power.larc.nasa.gov/data-access-viewer/} for the location with latitude 41.2087$^{\circ}$N and longitude 16.4016$^{\circ}$E. These are the coordinates of an existing large wind power plant in Italy. The GHI, i.e., the surface solar downward irradiation integrated over 285 to 2800 nm available at ground level, on a horizontal surface, and the clear-sky GHI, corresponding to cloudless weather conditions, were downloaded from \cite{cams2022gridded} at a temporal resolution of 15 min, for the location with latitude 42.5$^{\circ}$N and longitude 11.0$^{\circ}$E. These are the coordinates of an existing large PV power plant in Italy. 
Risk-free interest rates EONIA and €STR were downloaded from Refinitiv Datastream.

\subsection{Electricity price}
\label{sec:electricity_price_data}

\begin{figure}[H]
    \centering
    \includegraphics[width=.75\linewidth]{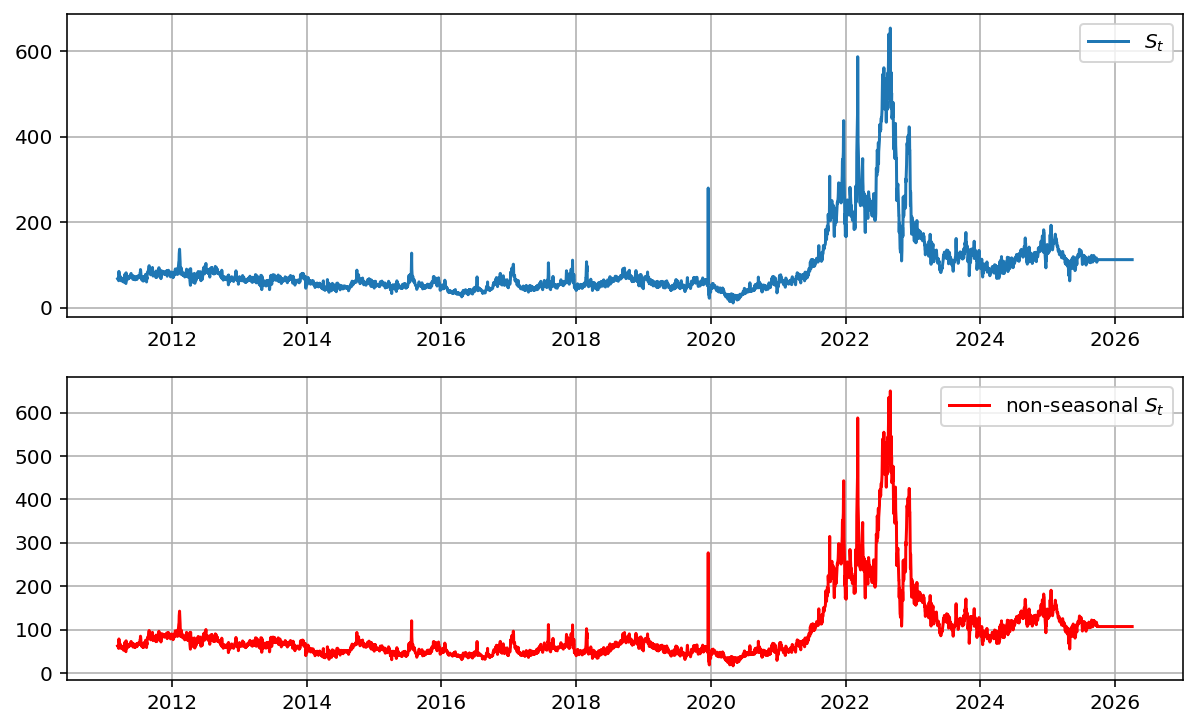}
    \caption{Electricity spot price (PUN)}
    \label{fig:pun_data_plot}
\end{figure}

Figure \ref{fig:pun_data_plot} shows the time series of the Italian spot prices of electricity (PUN) and the deseasonalized time series, while Table \ref{tab:electricity_parameters} holds the estimated model parameters. Seasonal effects can be seen to have limited impact on price, while the mean reverting structure of the price of electricity is visible in Figure \ref{fig:pun_data_plot} even in periods of high volatility and uncertainty, such as the beginning of 2022. More details on model selection can be found in Appendix \ref{sec:Electricity_price_preliminary_analysis}.

\begin{table}[H]
\centering
\caption{Estimated parameters of the electricity price model.}
\label{tab:electricity_parameters}
\begin{tabular}{lc}
\toprule
Parameter & Estimate \\
\midrule
$\alpha$ & 0.018719 \\
$\bar{X}$ & 35.082029 \\
$a_1^S$ & 4.403265 \\
$b_1^S$ & 5.766216 \\
$\beta$ & 0.000999 \\
$\bar{\nu}$ & 175.603123 \\
$\eta$ & 20.084123 \\
$\rho_{x,\nu}$ & 0.002734 \\
\bottomrule
\end{tabular}
\end{table}

\newpage

\subsection{Wind speed}
\label{sec:wind_data}

\begin{figure}[H]
    \centering
    \includegraphics[width=0.75\linewidth]{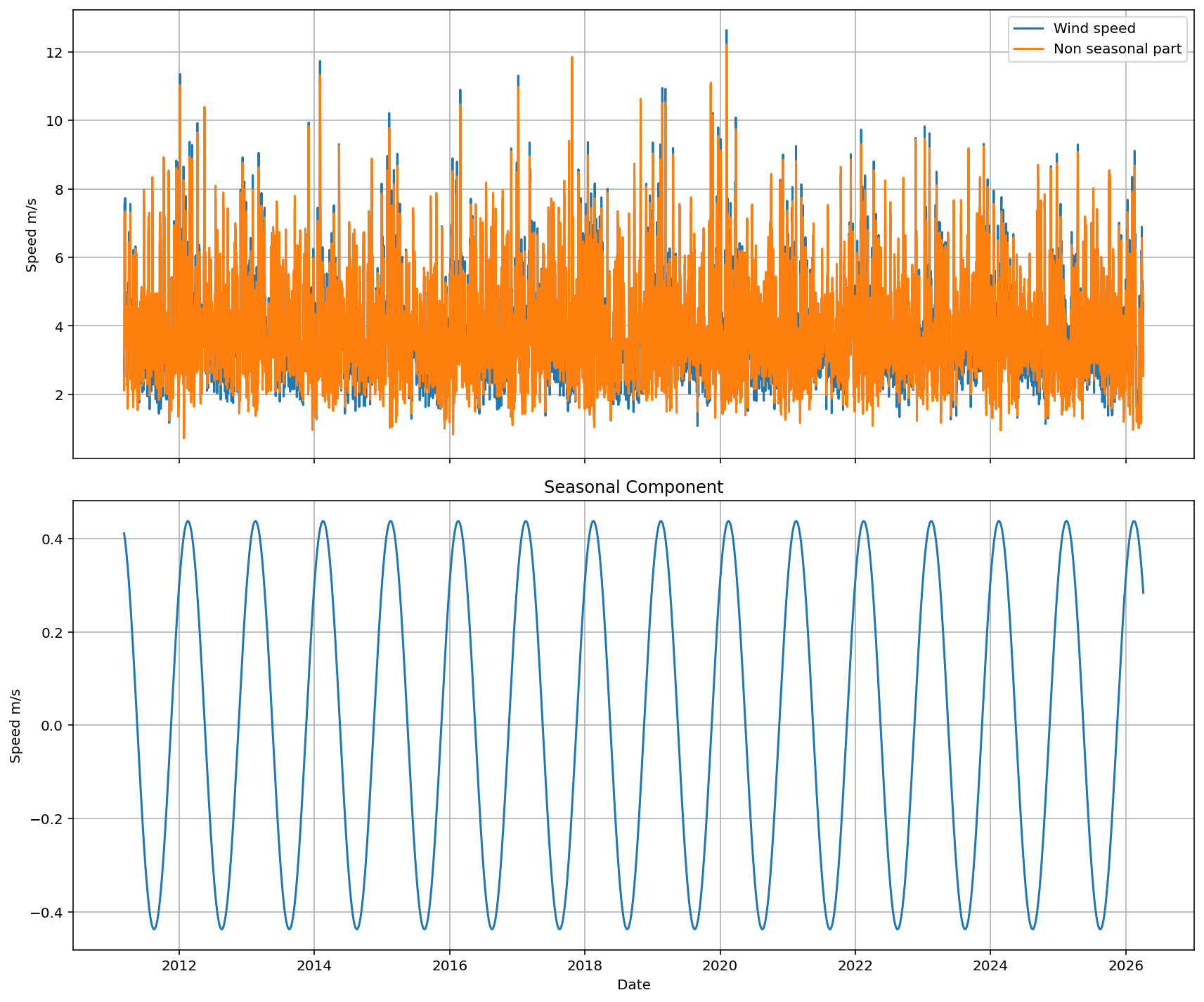}
    \caption{Wind Speed}
    \label{fig:wind_speed_data_plot}
\end{figure}

Figure \ref{fig:wind_speed_data_plot} shows the time series of average daily wind speed and its seasonal component. Its main features are positivity, seasonality, and mean-reversion. More details on model selection can be found in Appendix \ref{sec:Wind_preliminary_analysis}. The estimated model parameters are reported in Table \ref{tab:wind_parameters}.

\begin{table}[H]
\centering
\caption{Estimated parameters of the wind speed model.}
\label{tab:wind_parameters}
\begin{tabular}{lc}
\toprule
Parameter & Estimate \\
\midrule
$\kappa$ & 0.524054 \\
$\bar{Y}$ ($\theta$) & 3.837533 \\
$\sigma_y$ & 1.353790 \\
$a^{W}_{1}$ & -0.149610 \\
$b^{W}_{1}$ & 0.411152 \\
$\rho_{x,y}$ & -0.120000 \\
\bottomrule
\end{tabular}
\end{table}

\newpage

\subsection{Solar irradiance (GHI)}
\label{sec:solar_irradiance_data}
We provide a detailed explanation and justification of the modeling choice for GHI, as it constitutes the fourth main contribution of the present work.

\begin{figure}[H]
    \centering
    \includegraphics[width=0.97\linewidth]{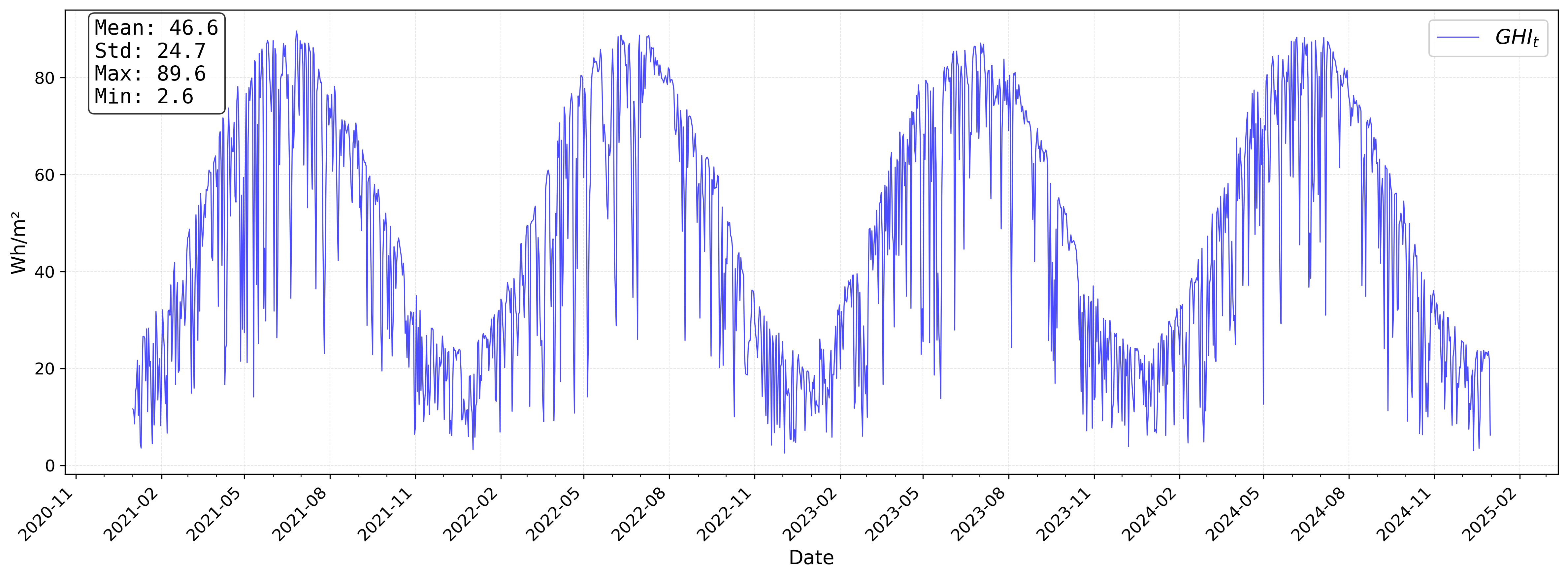}
    \caption{Average daily Global Horizontal Irradiance}
    \label{fig:ghi_daily}
\end{figure}

\begin{figure}[H]
    \centering
    \includegraphics[width=0.97\linewidth]{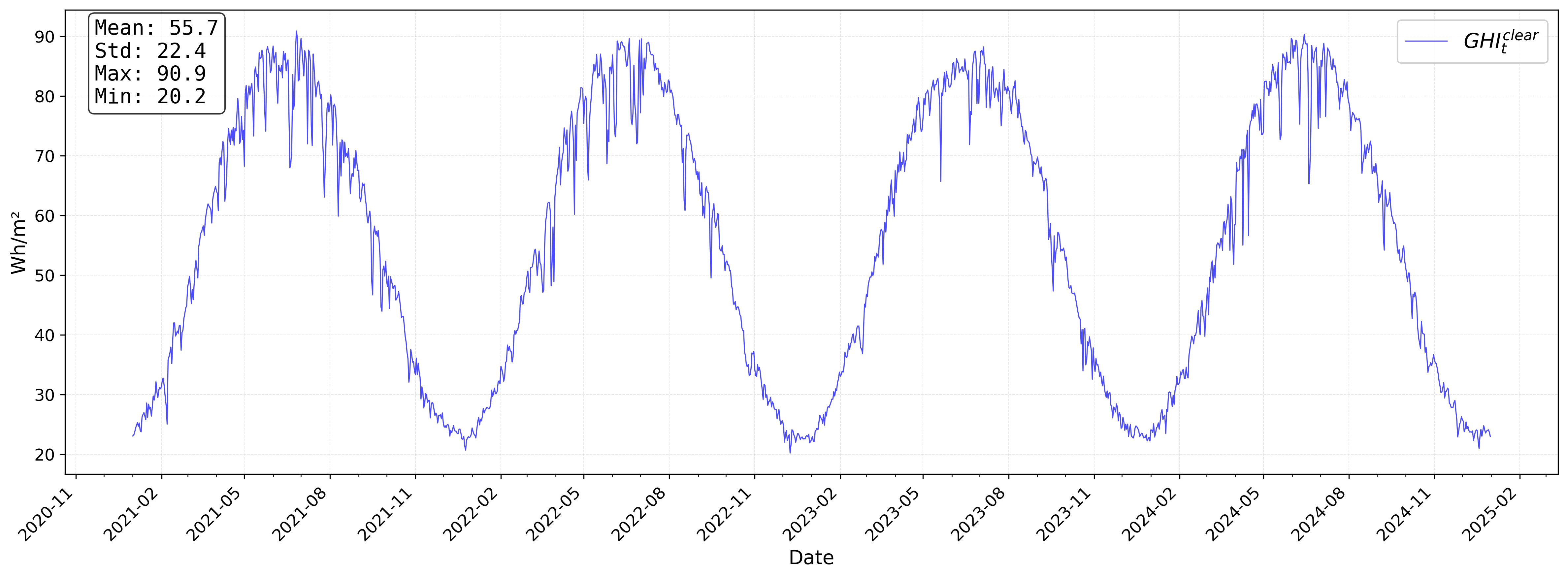}
    \caption{Average daily Clear-sky Global Horizontal Irradiance}
    \label{fig:clear_ghi_daily}
\end{figure}

The time series of average daily solar irradiance $GHI_t$ is presented in Figure \ref{fig:ghi_daily}.
Its main features are a strong seasonal component and its strictly positive values. 
Initial attempts at removing the seasonality component from the $GHI_t$ via additive or multiplicative models are unsuccessful, as the resulting time series still display very strong seasonal patterns. Similarly, attempts at removing seasonality from the natural logarithm of $GHI_t$ are unsuccessful. The results of these attempts are reported in Appendix \ref{app:ghi_attempts}.
We then decide to make use of the time series of average daily clear-sky irradiance, $GHI_t^{clear}$, provided by \cite{cams2022gridded}, estimated on the basis of the meteorological model of \cite{gschwind2019improving}, and presented in Figure \ref{fig:clear_ghi_daily}. 
We notice that the ratio $GHI_t/GHI_t^{clear}$ has a greatly reduced seasonal pattern. We therefore look for a deterministic approximation of $GHI_t^{clear}$, in order to maintain model tractability.
The work by \cite{reno2012global} collects and compares the performance of the main models for clear-sky irradiance introduced in the literature. Some of them are stochastic, depending on a number of additional weather variables, while others are deterministic and require only the zenith angle of the sun. The authors find that the deterministic model of clear-sky irradiance in \cite{haurwitz1945insolation} and \cite{haurwitz1946insolation} has the best performance among all of the deterministic models in the literature and is less accurate but comparable to the alternative stochastic models. It only depends on sun position, more precisely on the zenith angle $\theta_t$, which is a function of time and of the latitude, longitude, and altitude of the location in which the irradiance is measured.
This function has the form
\begin{equation}\label{eq:haurwitz}
    GHI_{\epsilon_t}^{H,clear} = 1098 \times cos(\theta_{\epsilon_t}) \times \exp\Big(-\frac{0.057}{cos(\theta_{\epsilon_t})}\Big),
\end{equation}
where the constants $1098$ and $0.057$ are fixed and taken from \cite{haurwitz1945insolation}.
It outputs values with a unit of measure of  W/m$^2$, which correspond to instantaneous values of solar irradiance from the Haurwitz model, which we therefore call $GHI_{\epsilon_t}^{H,clear}$. These values need to be converted into Wh/m$^2$ and aggregated daily, in order to recover the $GHI_{t}^{H,clear}$, which are in the same unit of measure of $GHI_t$.

\begin{figure}[H]
    \centering
    \includegraphics[width=0.97\linewidth]{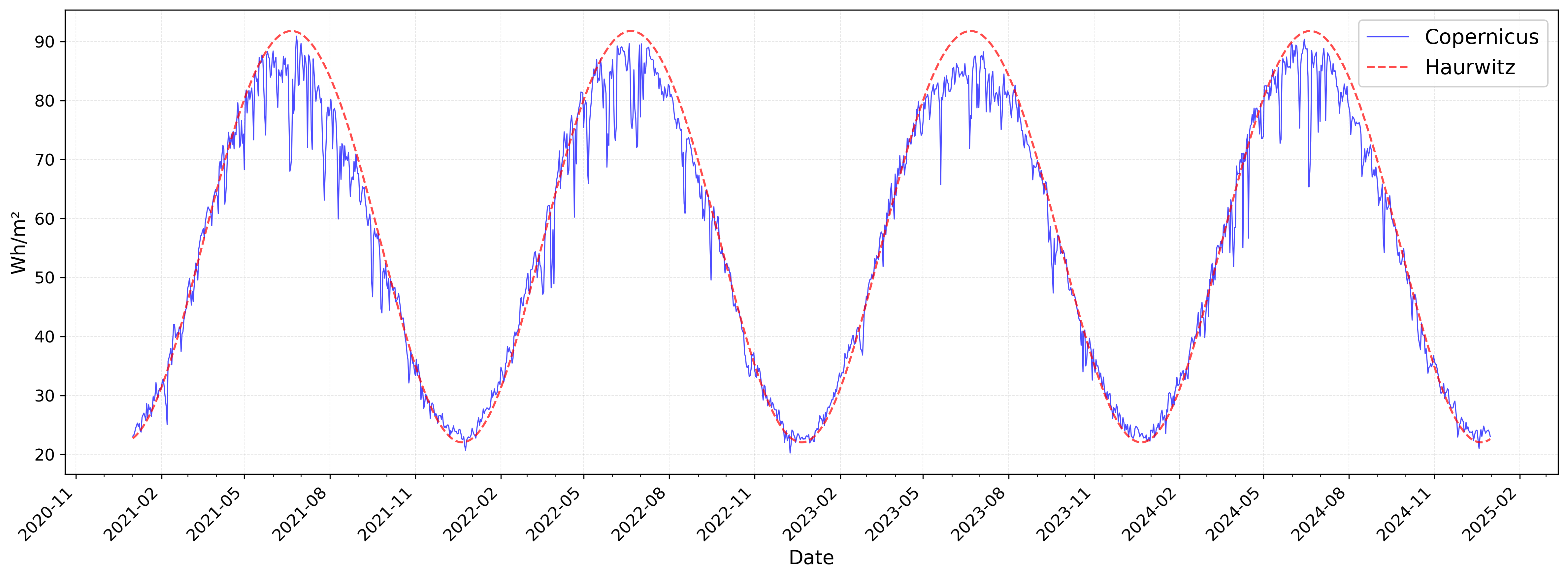}
    \caption{Haurwitz-model Clear-sky Global Horizontal Irradiance}
    \label{fig:haurwitz_clear_ghi_daily}
\end{figure}

We compute solar zenith angles using the \textit{pvlib} Python library based on the latitude (42.5$^{\circ}$N), longitude (11.0$^{\circ}$E), and altitude (133 m) of the selected location. Instantaneous irradiance values (W/m$^2$) are calculated using the Haurwitz formula for zenith angles below 90$^{\circ}$, corresponding to the daytime. These instantaneous values are converted to energy yields (Wh/m$^2$) by multiplying by 0.25 hours, assuming each 15-minute interval is represented by its midpoint. The minute length of the interval is selected to match that of the $GHI_t$ and $GHI_{t}^{clear}$ provided by \cite{cams2022gridded}. Daily means are then computed by averaging the 15-minute values over each calendar day, resulting in a daily time series of clear-sky GHI estimates. The values of $GHI_{t}^{H,clear}$ from the \cite{haurwitz1945insolation} model are plotted in Figure \ref{fig:haurwitz_clear_ghi_daily} against the Copernicus time series $GHI_{t}^{clear}$.
As the figure shows, $GHI_{t}^{H,clear}$ does not always dominate $GHI_{t}^{clear}$. We however want to ensure that the process $Z_t$ in Eq. \eqref{eq:logit_part_solar} lies in $(0,1)$, so that its logit transform exists and can be modeled with an SDE. For this reason, we require an estimate of $GHI_{t}^{clear}$ that dominates it, and thus also dominates $GHI_{t}$. 

\begin{figure}[H]
    \centering
    \includegraphics[width=0.97\linewidth]{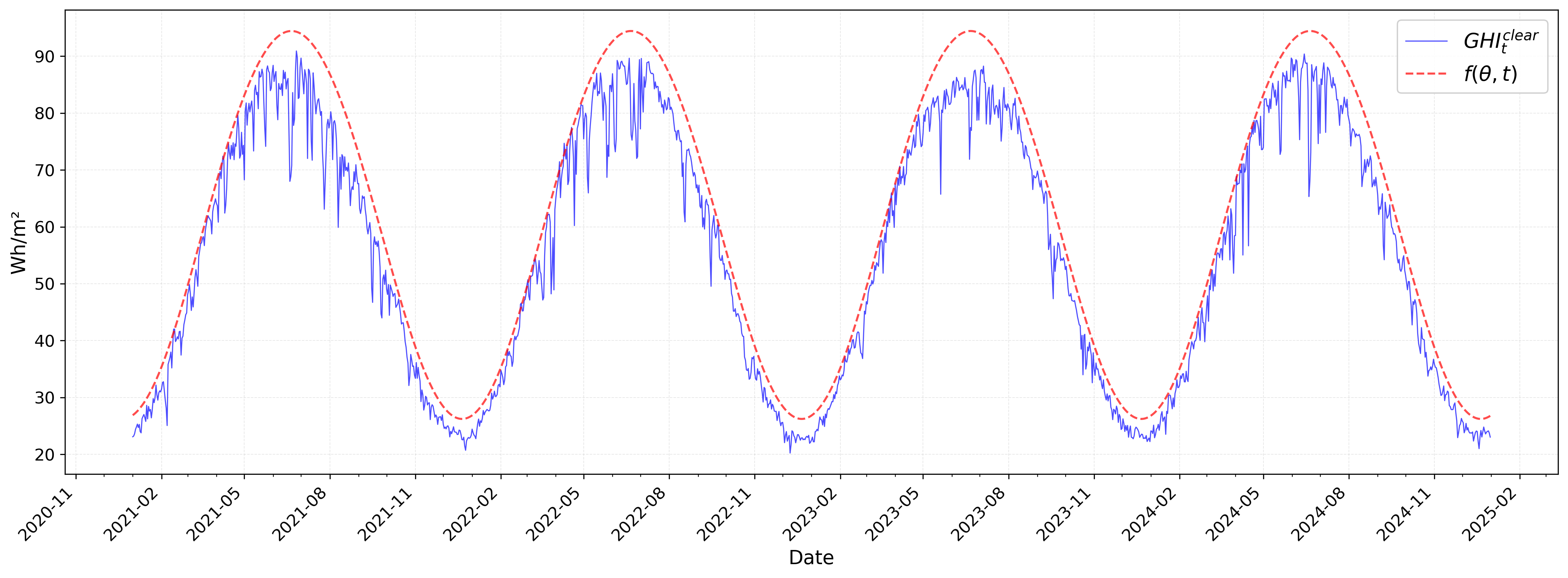}
    \caption{Envelope of Clear-sky GHI}
    \label{fig:envelope_clear_ghi_daily}
\end{figure}

\begin{figure}[H]
    \centering
    \includegraphics[width=0.97\linewidth]{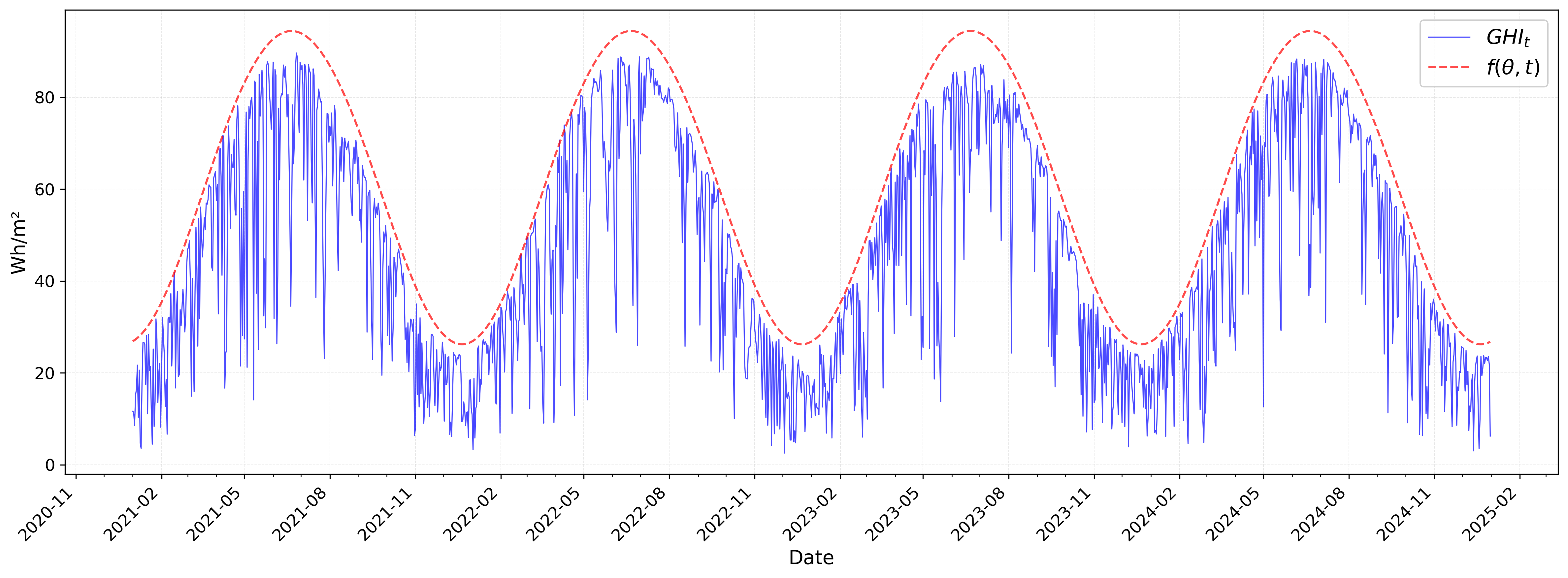}
    \caption{Envelope of Clear-sky GHI against daily GHI}
    \label{fig:envelope_ghi_daily}
\end{figure}

Therefore, the daily mean Haurwitz clear-sky GHI ($GHI_{t}^{H,clear}$ ) is finally linearly transformed to create an upper envelope for the Copernicus daily mean clear-sky GHI in the selected location. We denote this envelope as $f(\theta,t)$ and represent it as
\begin{equation}
f(\theta,t) = A \times GHI_{t}^{H,clear} + C,
\end{equation}
where $A = 0.972$ and $C = 1.23$ Wh/m$^2$ were optimized using quantile regression.
The envelope can now serve as an upper bound for $GHI_t$. It is presented in Figure \ref{fig:envelope_clear_ghi_daily} against the $GHI_t^{clear}$ and in Figure \ref{fig:envelope_ghi_daily} against the observed historical $GHI_t$.
The same linear transformation can be expressed directly in terms of the original instantaneous Haurwitz function $GHI_{\epsilon_t}^{H,clear}$. The daily mean clear-sky envelope then is
\begin{equation}
f(\theta,t) = A \times \frac{1}{n}\sum_{i=1}^{n} GHI_{\epsilon_i}^{H,clear} \times \Delta t + C
\end{equation}
where $\Delta t = 0.25$ hours (15-minute integration), and $n$ is the number of intervals per day. 

\begin{figure}[H]
    \centering
    \includegraphics[width=0.97\linewidth]{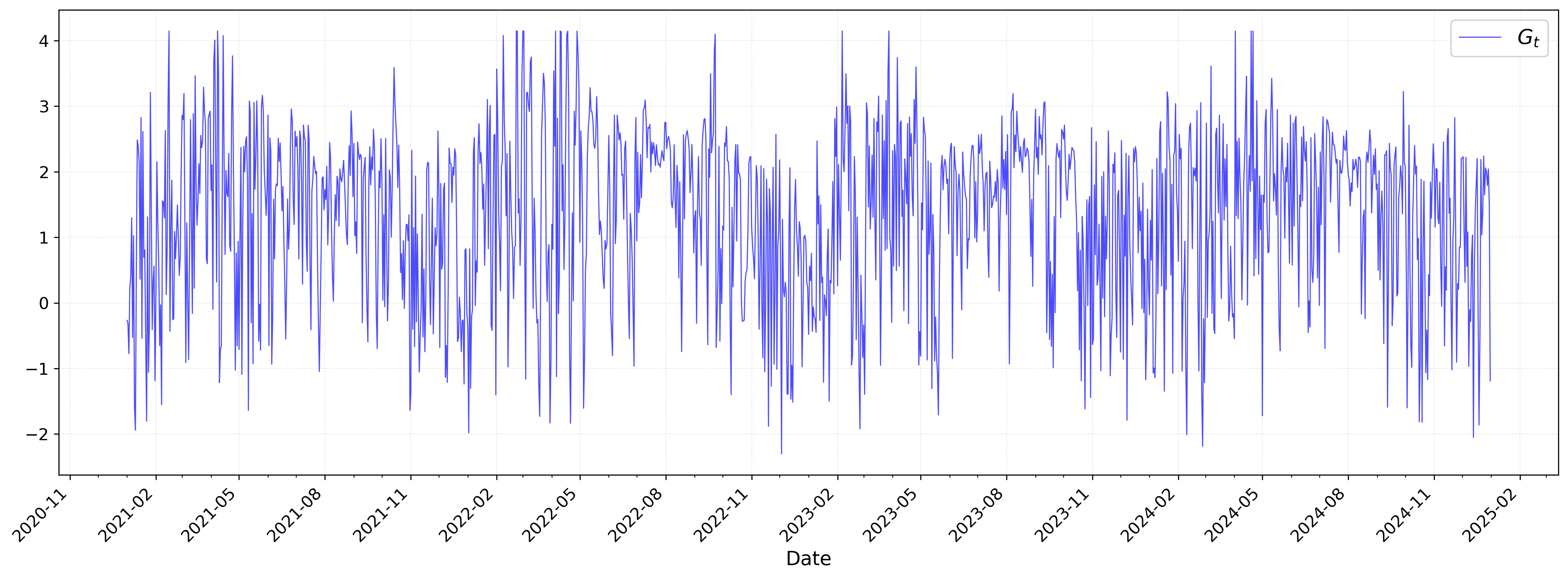}
    \caption{Time series of the extracted $G_t$}
    \label{fig:solar_increments}
\end{figure}

\begin{figure}[H]
    \centering
    \includegraphics[width=0.97\linewidth]{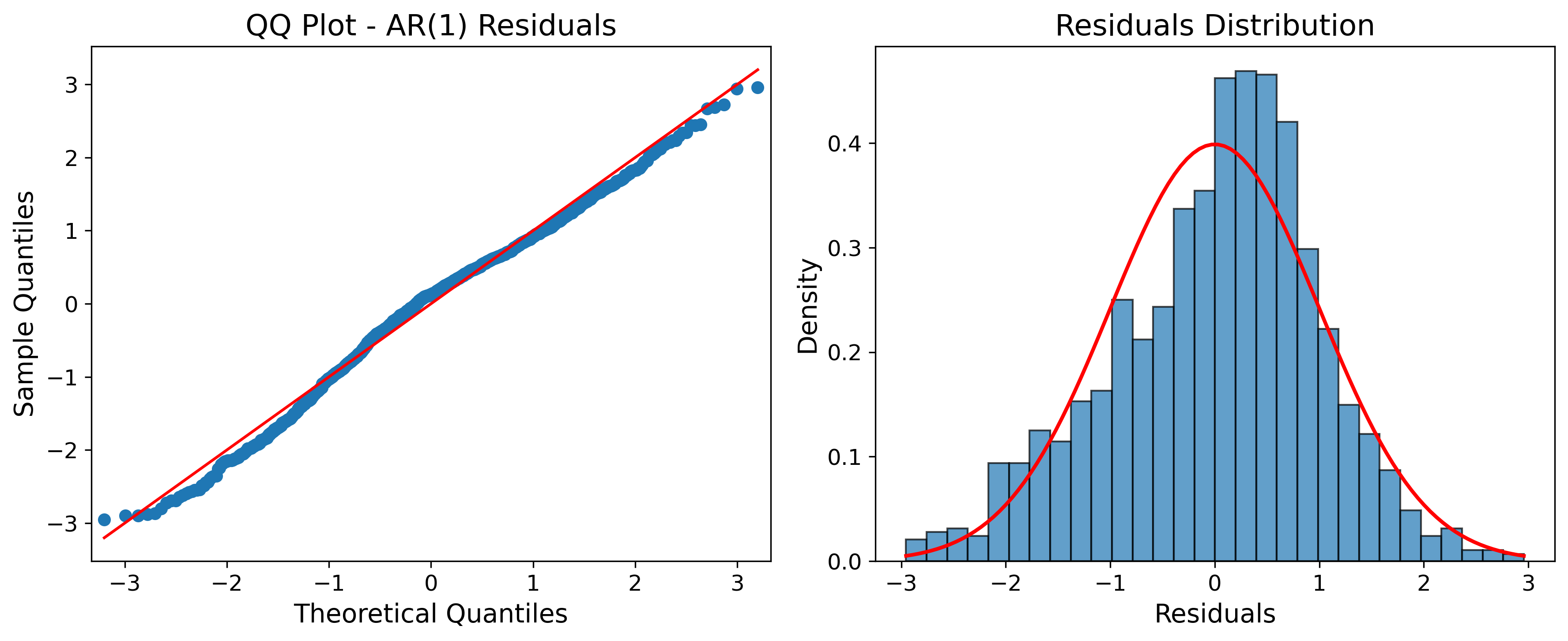}
    \caption{QQ plot and residuals distribution with Gaussian innovations}
    \label{fig:qq_gauss}
\end{figure}

\begin{figure}[H]
    \centering
    \includegraphics[width=0.97\linewidth]{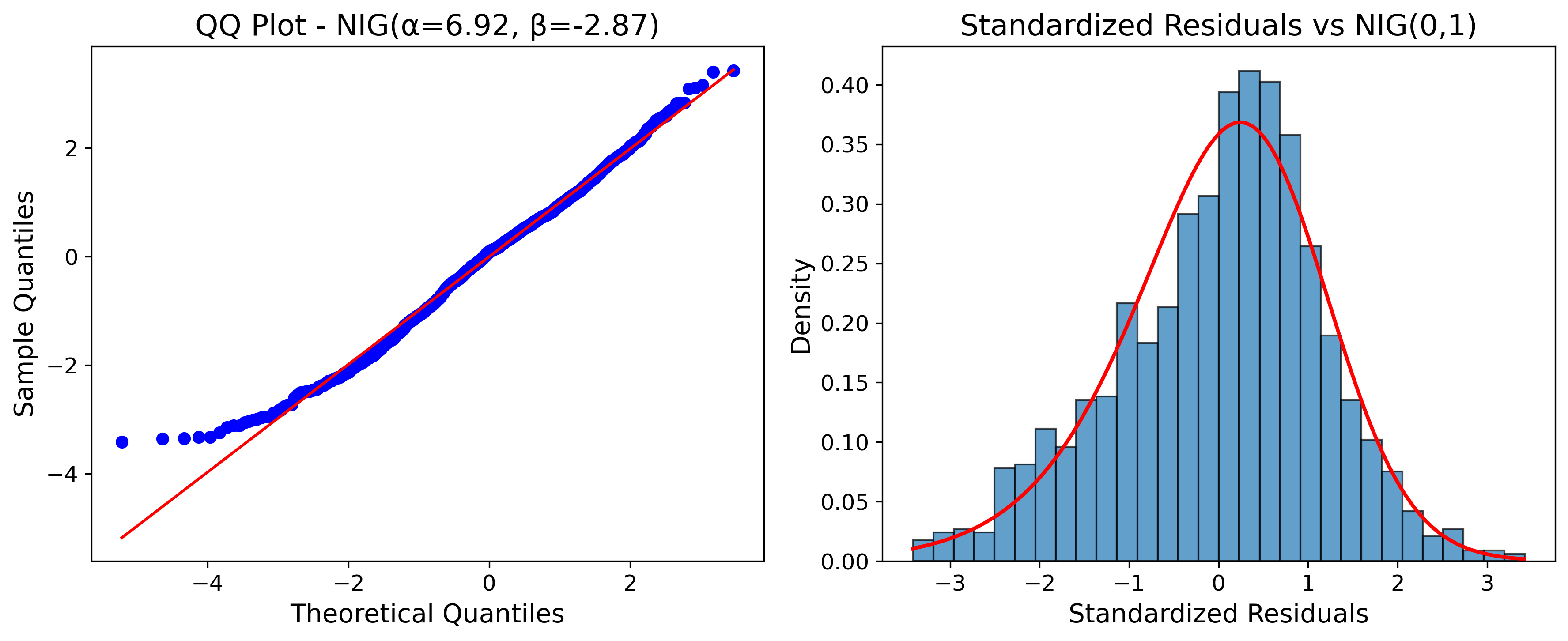}
    \caption{QQ plot and residuals distribution with NIG innovations}
    \label{fig:qq_nig}
\end{figure}

Now that $f(\theta,t)$ is calibrated, we compute the ratio $GHI_t/f(\theta,t)$ in Eq. \eqref{eq:solar_irradiance} to recover $Z_t$. We then apply the logit transform to $Z_t$ and obtain the series of the observed $\Lambda^G_t + G_t$. The seasonality function in Eq. \eqref{eq:seasonal_part_solar} is then calibrated and the $G_t$ are finally extracted. They are plotted in Figure \ref{fig:solar_increments}.
Based on their autocorrelation, we fit on the $G_t$ an auto-regressive model of order 1, first with Gaussian and then, due to their skewed nature, with Normal Inverse Gaussian (NIG) innovations. Figure \ref{fig:qq_gauss} holds residual plots of the Gaussian model, while Figure \ref{fig:qq_nig} holds them for the NIG model. Kolmogorov-Smirnov tests show unsatisfactory fits of both distributions, but since the Gaussian QQ plot achieves a better result in the tails of the distribution, we select an SDE with Gaussian increments and which presents mean-reversion (to capture the auto-regressive nature of the data).\\

\begin{figure}[H]
    \centering
    \includegraphics[width=0.97\linewidth]{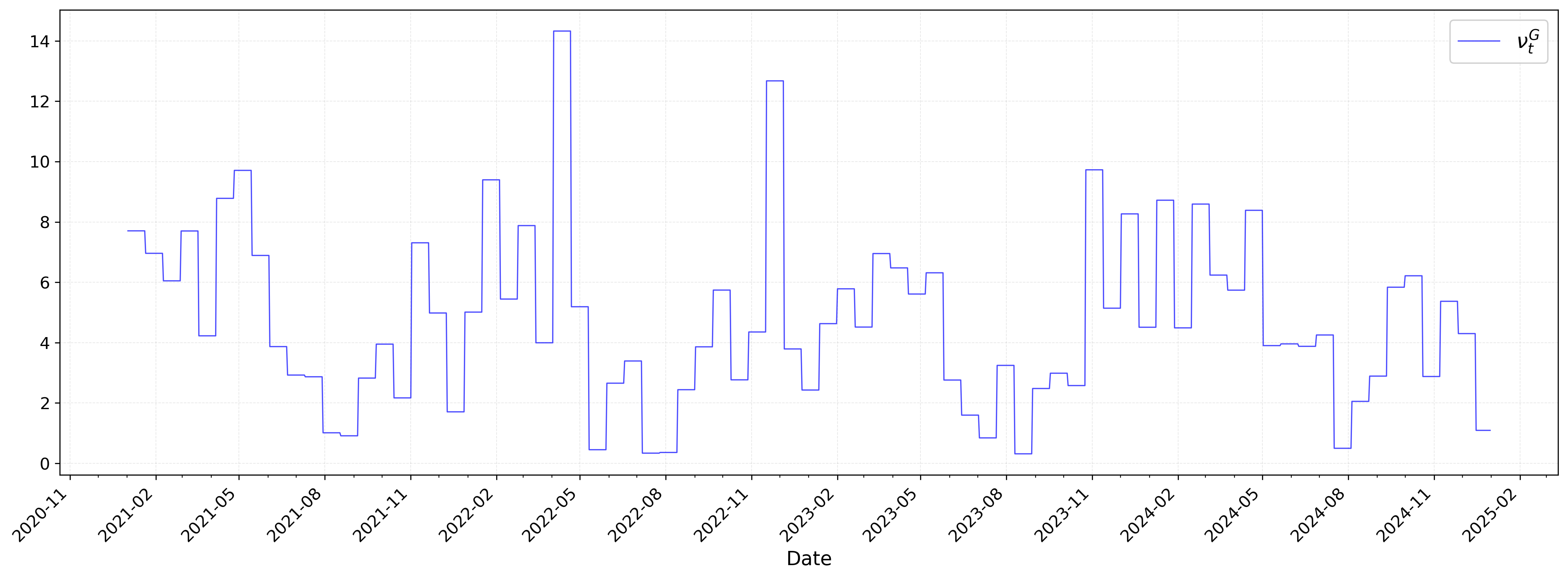}
    \caption{Estimated latent variance process}
    \label{fig:stoch_vol_ghi}
\end{figure}

The imperfect alignment with the Gaussian distribution could be caused by stochastic volatility. To test for its presence, we apply the Ljung-Box test to the squared residuals. The null hypothesis of no autocorrelation is tested at lags 10, 20, and 30. All test p-values are below the 0.01 significance level and we hence reject the null hypothesis of no autocorrelation. 
This result provides strong evidence of serial dependence in the squared residuals, suggesting the presence of stochastic volatility in the time series. 
Consequently, our final model for $G_t$ is the one in Eq. \eqref{eq:ou_part_solar} and \eqref{eq:volatility_solar}, which involves mean-reversion, Gaussian increments, and stochastic volatility. We calibrate it on the extracted $G_t$ with the procedure detailed in \cite{alfonsi2024stochastic}. The estimated latent variance process is computed for buckets of 19 days and is shown in Figure \ref{fig:stoch_vol_ghi}. The number of days is selected empirically in order to ensure the numerical success of the OLS-based estimation procedure. The parameter estimates are reported in Table \ref{tab:estimates_ghi}.

\begin{table}[H]
    \centering
\begin{tabular}{lc}
    \toprule
    Parameter & Estimate \\
    \midrule
    $\alpha^G$ & 1.108047 \\
    $\bar{G}$ & 1.326537 \\
    $a^{G}_{1}$ & 0\\
    $b^{G}_{1}$ & -0.474562 \\
    $\beta^G$ & 0.00124 \\
    $\bar{\nu}^G$ & 4.535051 \\
    $\eta^G$ & 0.391538 \\
    $\rho_{g,\nu^{G}}$ & -0.072367\\
    $\rho_{g,x}$ & 0.020175 \\
    \bottomrule
\end{tabular}
    \caption{Estimated parameters of the GHI model}
    \label{tab:estimates_ghi}
\end{table}

\newpage

\section{Empirical Analysis}
\label{sec:empirical_analysis}
In this section we perform a numerical illustration of the risk-neutral pricing of the types PPAs defined in Section \ref{sec:contract_definitions_and_risk_neutral_pricing}, and of their risk-assessment, as described in Section \ref{sec:risk_assessment}. Since we do not have access to the time series of strike prices of real contracts in one singular location, we rely on Monte Carlo methods to determine them. The following simulations assume a delivery period of one day and a settlement period of one month. The risk-free discount factors are computed based on the values of EONIA and, when its time series stops being available, of €STR. On every valuation date, the quoted interest rate term structures of that same day are used to discount the projected future cash flows.


\subsection{Fixed-price PPA risk-neutral price}
\begin{figure}[H]
    \centering
    \includegraphics[width=0.75\linewidth]{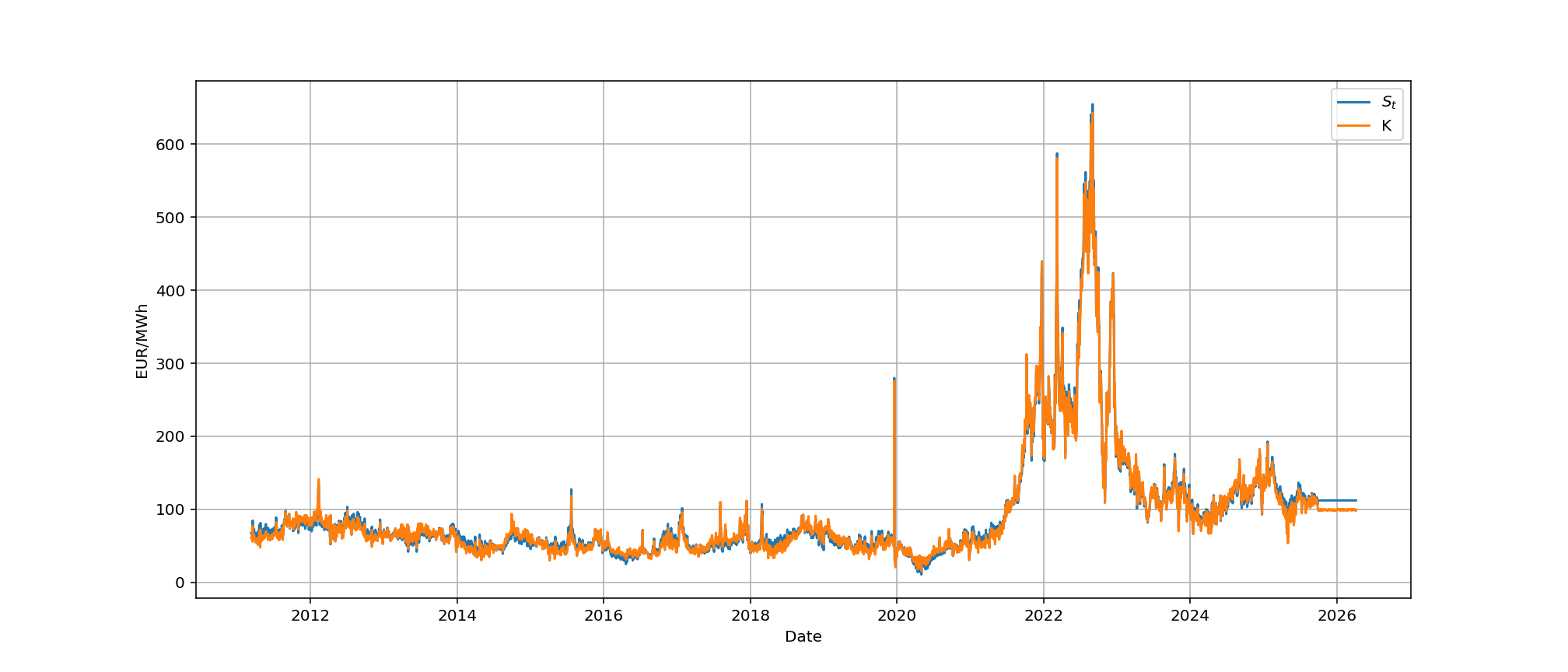}
    \caption{PPA price of a fixed-price contract for a wind power producer}
    \label{fig:ppa_price_plain_vanilla_wind}
\end{figure}

\begin{figure}[H]
    \centering
    \includegraphics[width=0.75\linewidth]{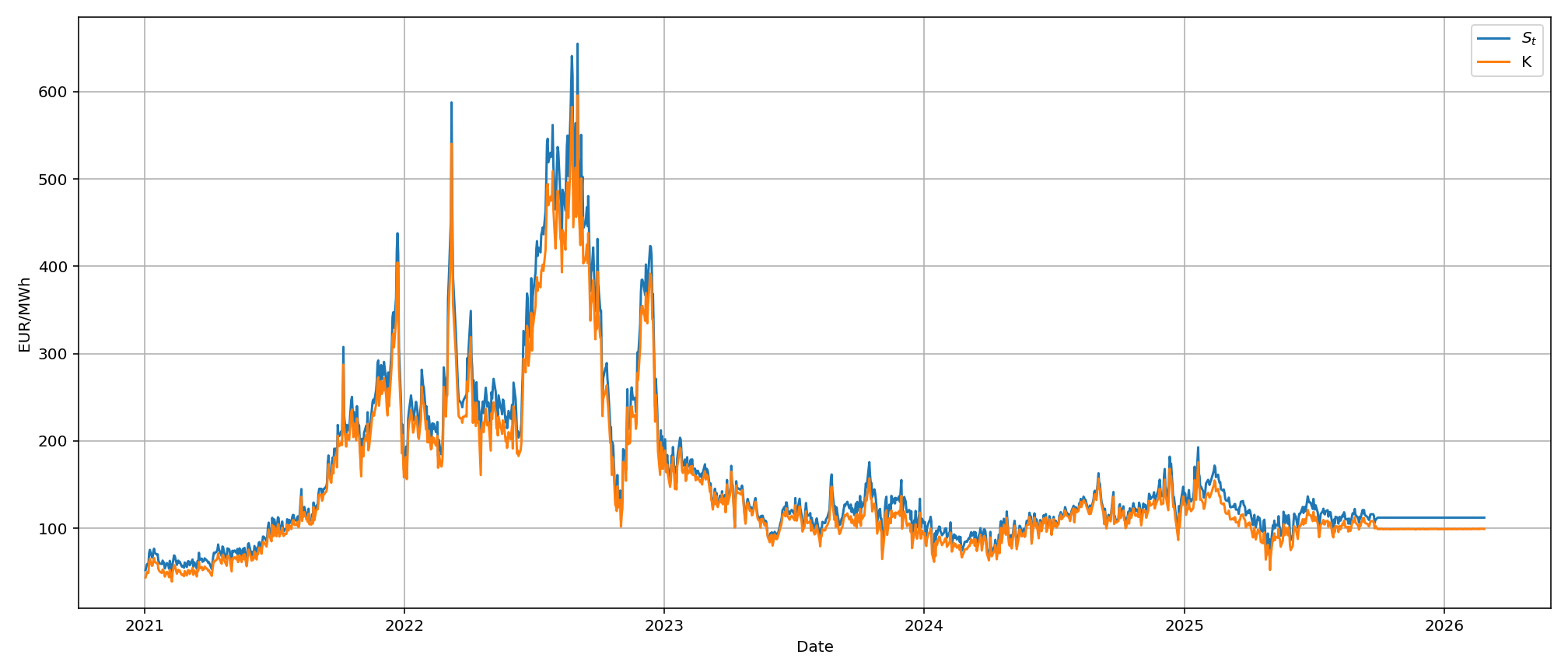}
    \caption{PPA price of a fixed-price contract for a solar power producer}
    \label{fig:ppa_price_plain_vanilla_solar}
\end{figure}

Figures \ref{fig:ppa_price_plain_vanilla_wind} and \ref{fig:ppa_price_plain_vanilla_solar} show the time series of risk-neutral prices of wind and PV PPAs, respectively, computed each day for a contract having a tenor of 10 years, solving Eq. \eqref{eq:ppa_fair_price_generic}, based on the updated information available on that day. 
The risk-neutral prices are plotted together with the observed spot electricity prices (PUN). We note that, even though the risk-neutral prices of both wind and PV PPAs follow the spot price of electricity, they are both lower than it.

\begin{figure}[H]
    \centering
    \includegraphics[width=0.75\linewidth]{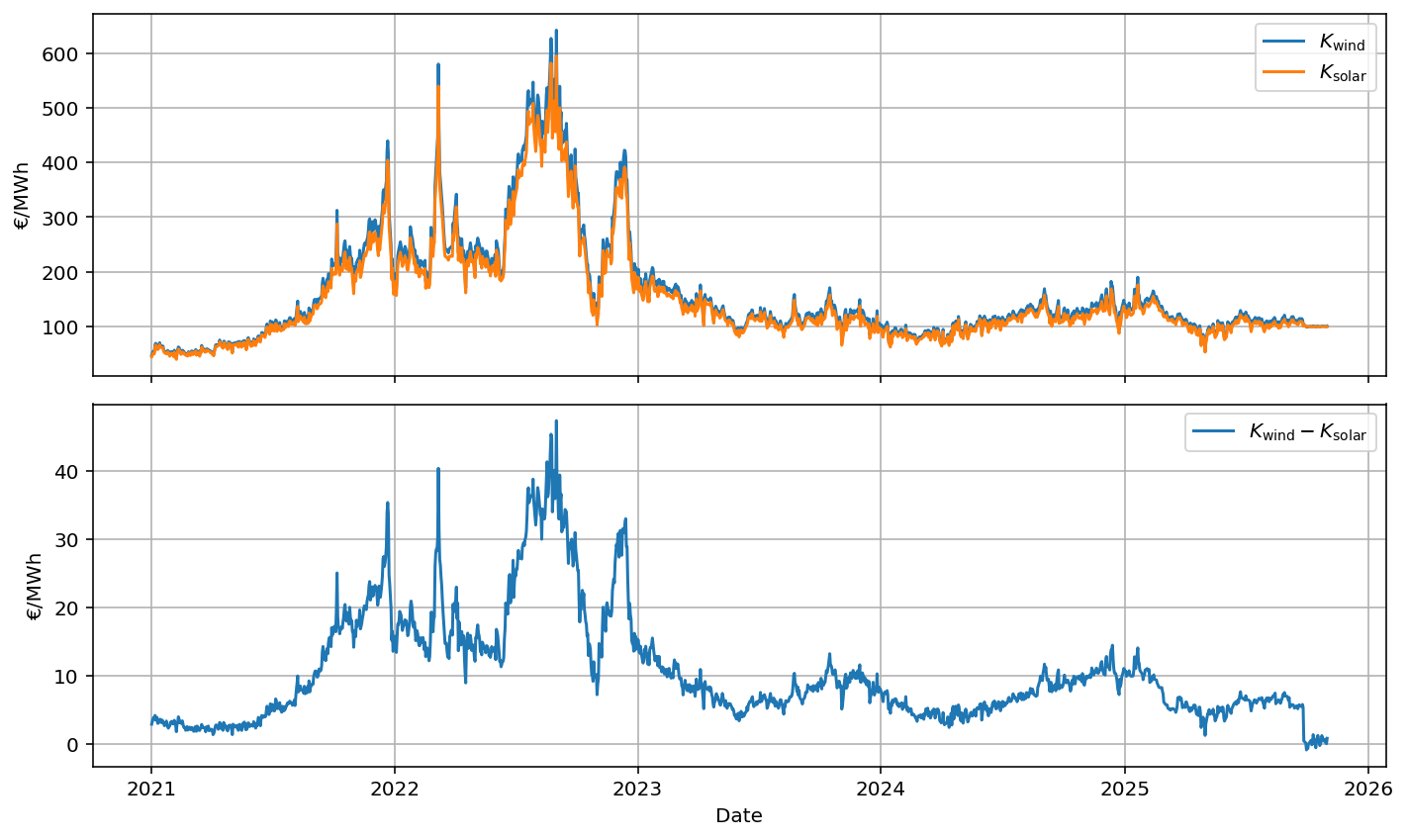}
    \caption{Difference between fair price of a wind PPA and solar PPA}
    \label{fig:plain_vanilla_wind_vs_solar}
\end{figure}

Figure \ref{fig:plain_vanilla_wind_vs_solar} shows that the risk-neutral price for wind power PPAs is usually higher than the one of PV PPAs and that such difference is greater in periods of high volatility and in the presence of market shocks. This is also confirmed by a statistical t-test performed on the difference between the two prices (see Table \ref{tab:t_test_comparison}), which results in the rejection of the null hypothesis $H_0: \mathbb{E}[K_{wind} - K_{PV}] =0$.

\subsection{Stepped PPA risk-neutral price}
\begin{figure}[H]
    \centering
    \includegraphics[width=0.75\linewidth]{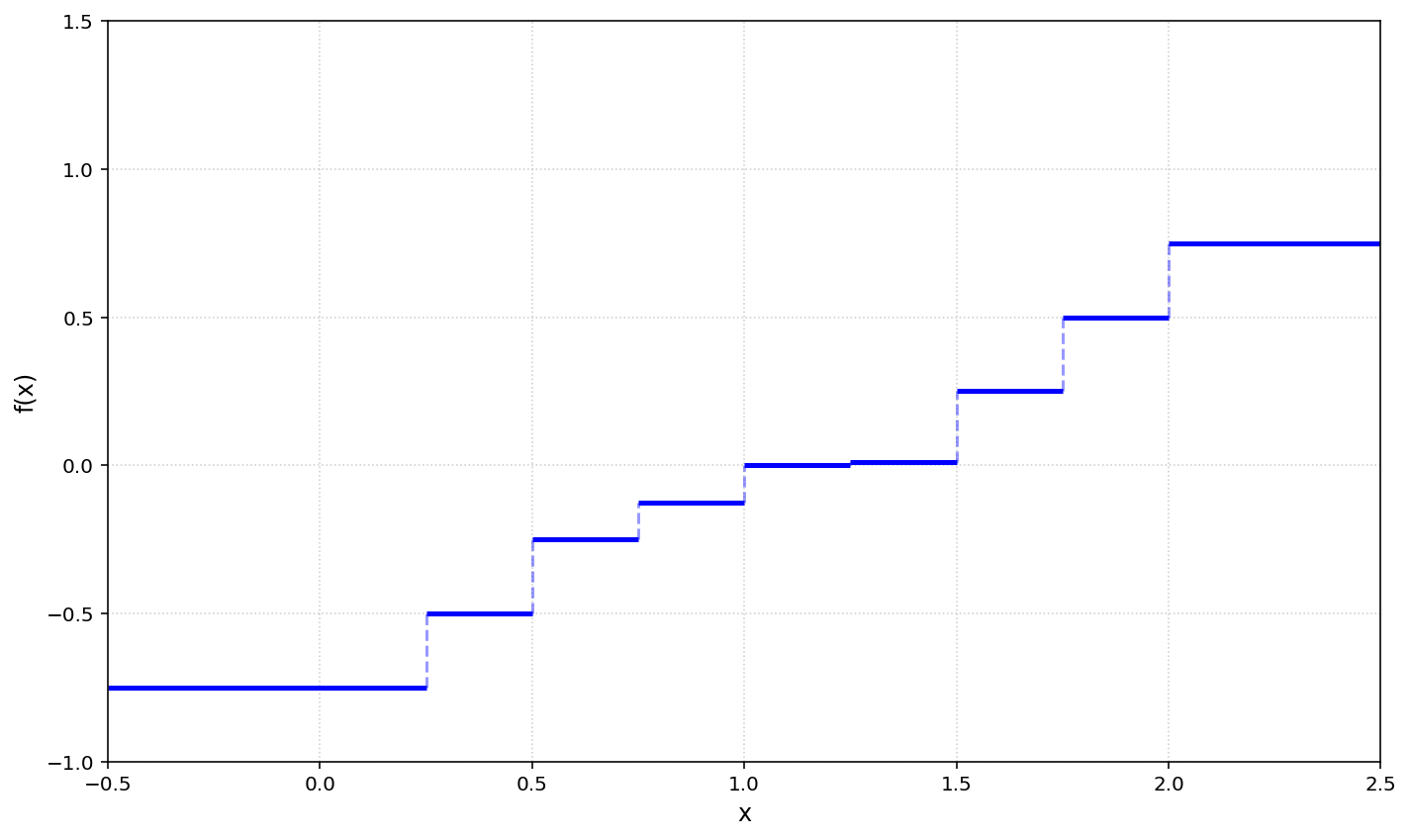}
    \caption{Step function of stepped PPA}
    \label{fig:step_function}
\end{figure}

\begin{figure}[H]
    \centering
    \includegraphics[width=0.75\linewidth]{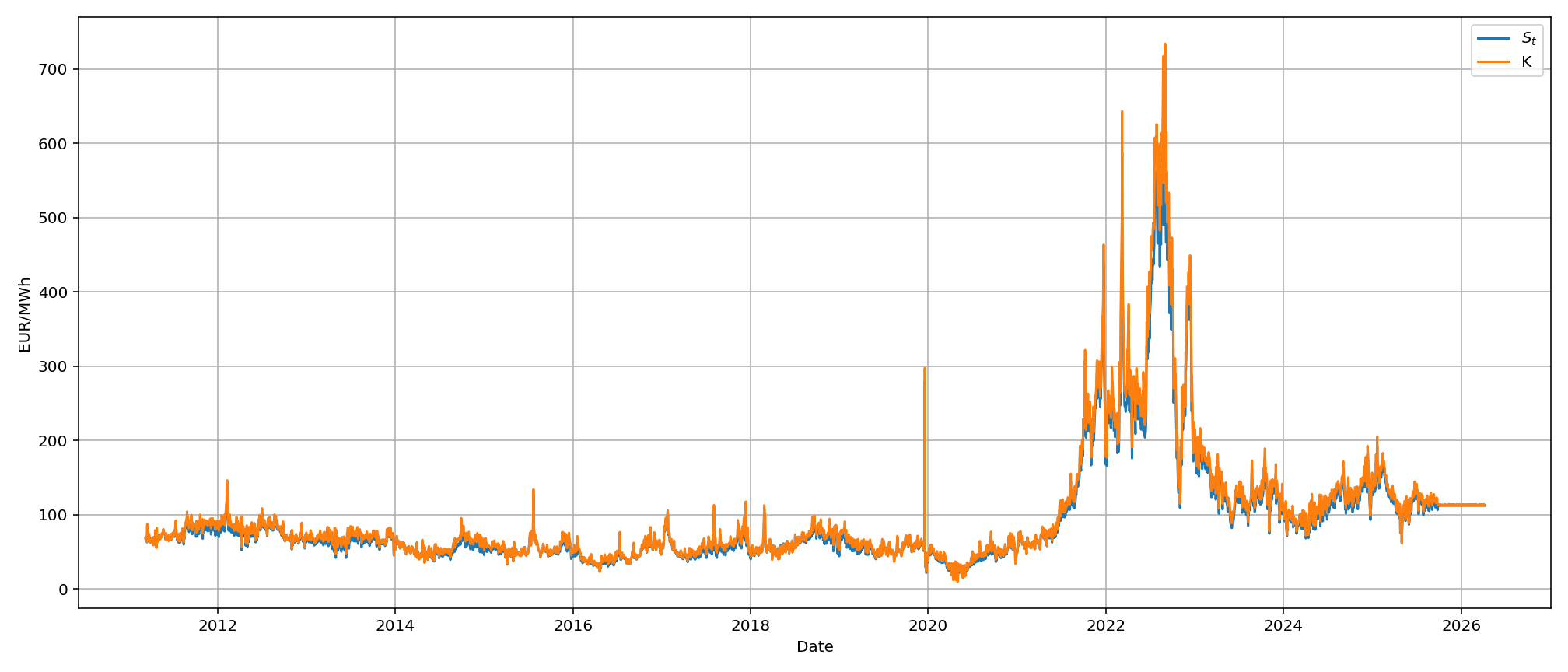}
    \caption{PPA price of a stepped ppa contract for a wind power producer}
    \label{fig:ppa_price_stepped_conytract_wind}
\end{figure}

\begin{figure}[H]
    \centering
    \includegraphics[width=0.75\linewidth]{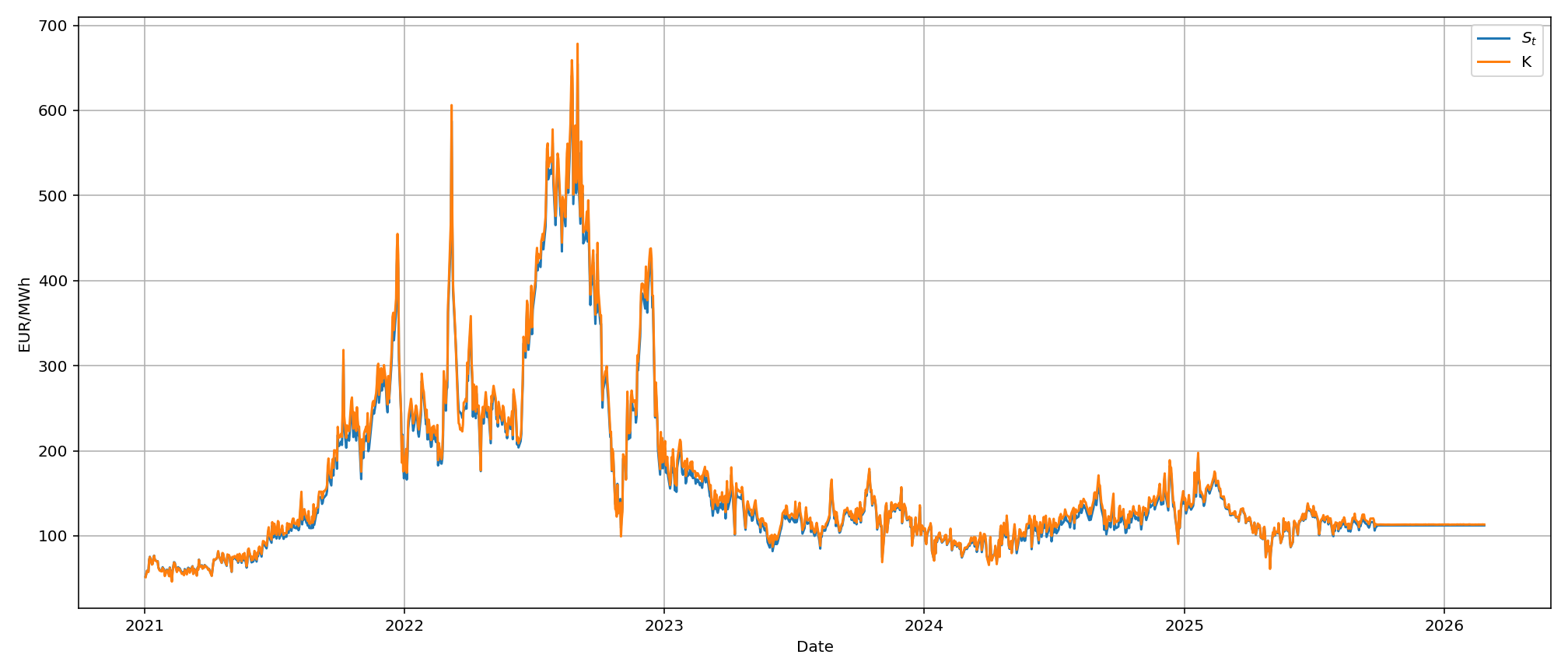}
    \caption{PPA price of a stepped ppa contract of a solar power producer}
    \label{fig:ppa_price_stepped_conytract_solar}
\end{figure}

Figures \ref{fig:ppa_price_stepped_conytract_wind} and \ref{fig:ppa_price_stepped_conytract_solar} display the time series of risk-neutral prices of wind and PV PPAs, respectively, computed each day for a contract having a tenor of 10 years, solving Eq. \eqref{eq:stepped_ppa_price}, based on the updated information available on that day. The selected shape of the step function $f(S,T)$ is shown in Figure \ref{fig:step_function}. The values on the x-axis represent the ratio between the spot electricity price in a delivery period, $S(T_j)$, and the spot electricity price on the valuation date, $S(0)$. The contractual terms make it so that the price paid by the offtaker in each delivery period $T_j$, $K(1+f(S,T_j))$, increases or decreases with respect to $K$, depending on the value of $x = S(T_j)/S(0)$. 
The risk-neutral prices are plotted together with the observed spot electricity prices (PUN). We again note that, even though the risk-neutral prices of both wind and PV PPAs follow the spot price of electricity, they are both lower than it.

\begin{figure}[H]
    \centering
    \includegraphics[width=0.75\linewidth]{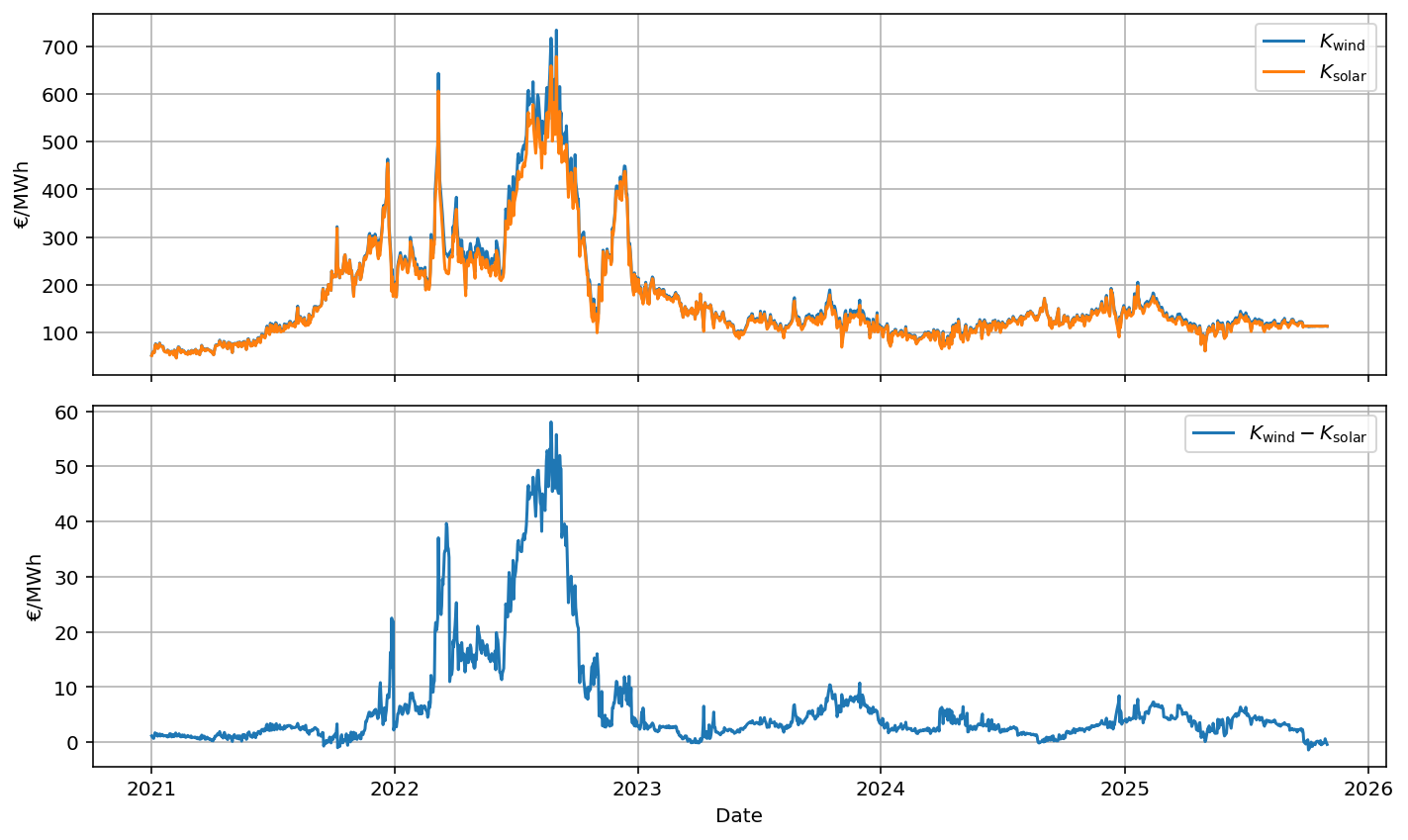}
    \caption{Difference between fair price of a wind stepped PPA and stepped collar PPA}
    \label{fig:stepped_ppa_wind_vs_solar}
\end{figure}

As before, Figure \ref{fig:stepped_ppa_wind_vs_solar} and Table \ref{tab:t_test_comparison} confirm again that the risk-neutral prices of PV PPAs are lower than those of the wind ones.

\begin{table}[htbp]
\centering
\caption{One-sample t-tests for the difference between wind and PV risk-neutral prices across PPA contract structures. The null hypothesis is $H_0:\mathbb{E}[K_{\mathrm{wind}}-K_{\mathrm{PV}}]=0$.}
\label{tab:t_test_comparison}
\begin{tabular}{lrrrrr}
\hline
Contract & Mean & Std. Dev. & t-statistic & p-value & Decision \\
\hline
Fixed-price & 10.6513 & 8.1935 & 53.2508 & $<10^{-16}$ & Reject $H_0$ \\
Reverse-Collar & 9.8492 & 7.6640 & 52.6432 & $<10^{-16}$ & Reject $H_0$ \\
Stepped & 6.6906 & 9.9937 & 27.4243 & $4.17\times10^{-137}$ & Reject $H_0$ \\
\hline
\end{tabular}
\end{table}
 \newpage

\subsection{Reverse Collar PPA risk-neutral price}
\begin{figure}[H]
    \centering
    \includegraphics[width=0.75\linewidth]{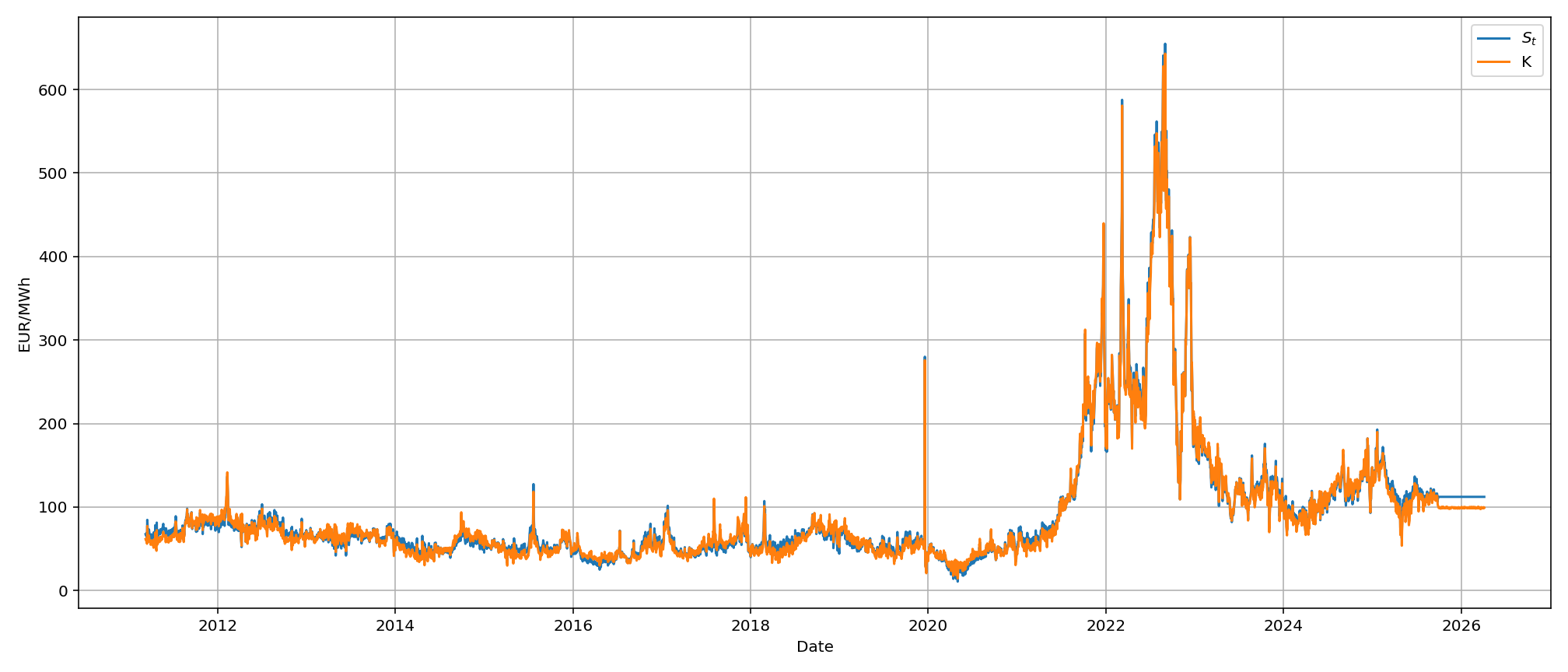}
    \caption{PPA price of a reverse collar contract for a wind power producer}
    \label{fig:ppa_price_reverse_collar_wind}
\end{figure}

\begin{figure}[H]
    \centering
    \includegraphics[width=0.85\linewidth]{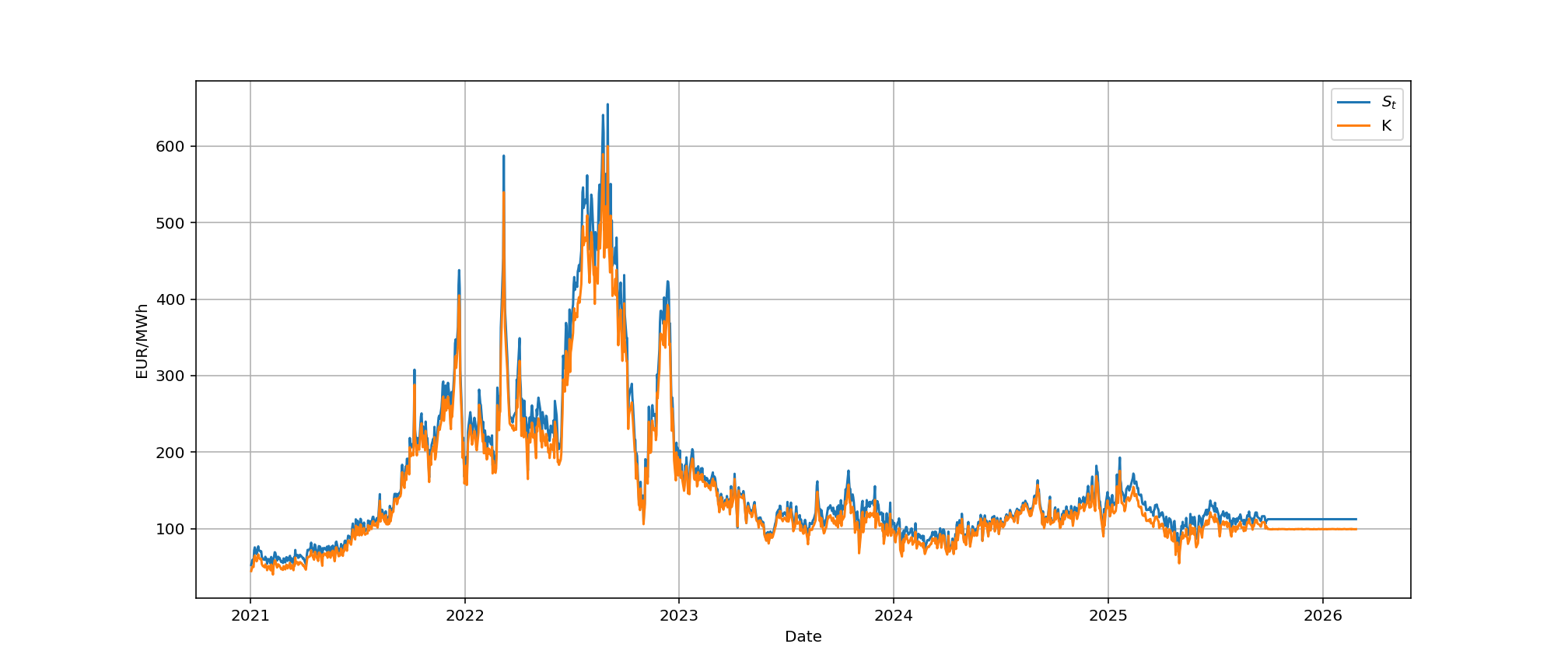}
    \caption{PPA price of a reverse collar contract for a solar power producer}
    \label{fig:ppa_price_reverse_collar_solar}
\end{figure}

Figures \ref{fig:ppa_price_reverse_collar_wind} and \ref{fig:ppa_price_reverse_collar_solar} show, for each day, the risk-neutral price with a tenor of 10 years for PPAs on wind and PV power, respectively, computed following Eq. \eqref{eq:fair_strike_collar_ppa}. On each valuation day, the contractual values of $K_{min}$ and $K_{max}$ are set to be equal to 70\% and 130\%, respectively, of the valuation day's observed spot electricity price.
Each fair $K$ is also compared to the observed PUN and even though both follow the price of electricity, they are again both lower than it.

As in the other cases, the fair PPA price is again lower than the observed spot electricity price, showing that, on a daily basis, given the current information, entering into a fair PPA contract would be cheaper than directly going on the market, (see Figures \ref{fig:ppa_price_reverse_collar_wind} and \ref{fig:ppa_price_reverse_collar_solar}). 

\begin{figure}[H]
    \centering
    \includegraphics[width=0.75\linewidth]{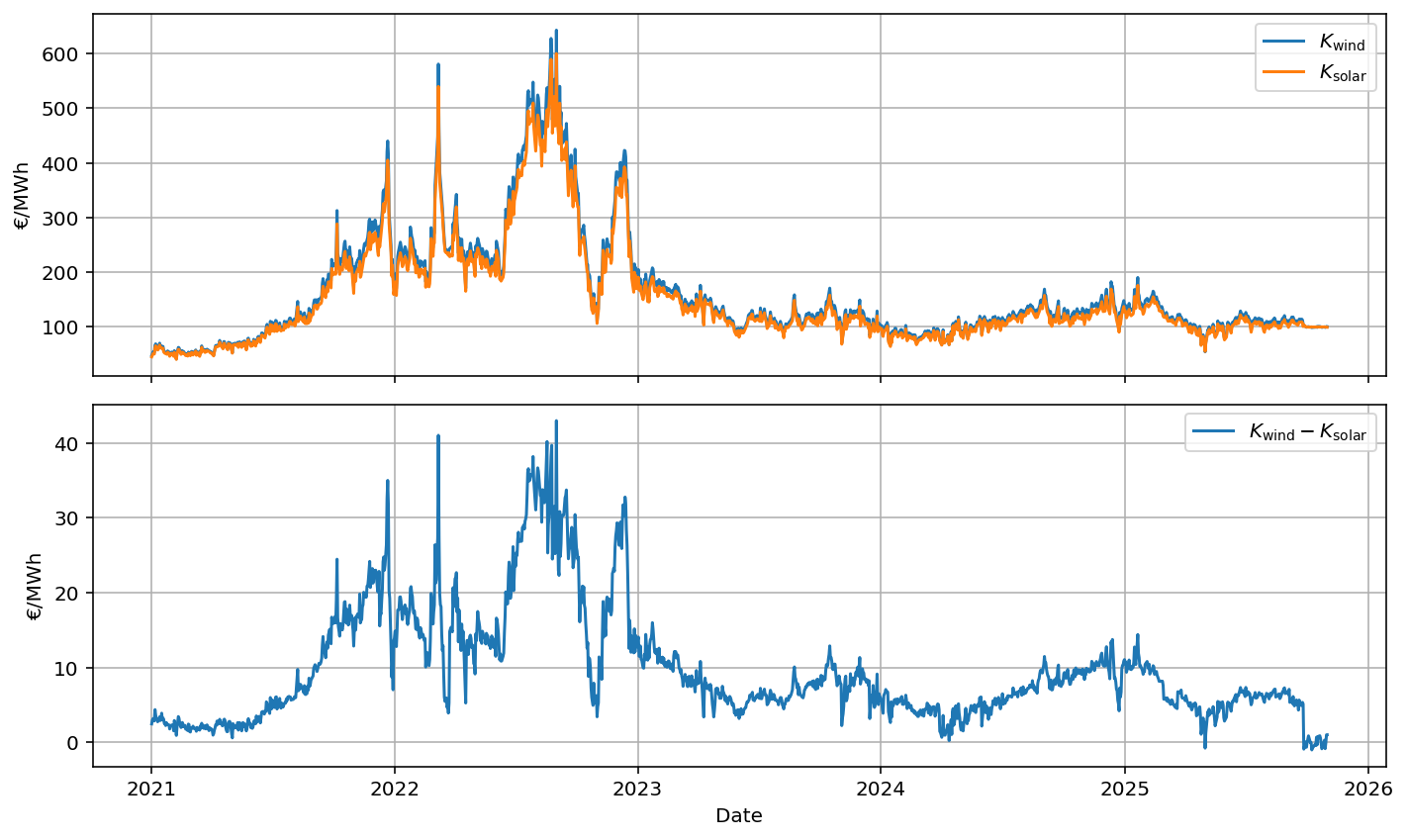}
    \caption{Difference between fair price for a wind reverse collar PPA and solar reverse collar PPA}
    \label{fig:reverse_collar_wind_vs_solar}
\end{figure}

Figure \ref{fig:reverse_collar_wind_vs_solar} and Table \ref{tab:t_test_comparison} confirm again that contracts on PV power have lower risk-neutral prices than wind ones.

\subsection{Payoff analysis and risk assessment}
\label{sec:payoff_analysis}
In this section we perform a numerical analysis of the  terminal contract payoff risk-assessment and valuation results, as described in Section \ref{sec:risk_assessment}.

The payoff distributions are reported in Tables \ref{tab:pv_solar_risk}, \ref{tab:rc_solar_risk}, \ref{tab:step_solar_risk}, \ref{tab:pv_wind_risk}, \ref{tab:rc_wind_risk}, and \ref{tab:step_wind_risk}. They reveal substantial differences in the risk-return characteristics of the considered PPA structures and technologies. 
Among the fixed-price contracts, the PV PPA exhibits a relatively symmetric payoff distribution, with low skewness and moderate dispersion, while the wind PPA is characterized by significantly higher volatility and much heavier tails. In particular, the standard deviation of the wind contract noticeably exceeds that of the solar contract, and the extremely high kurtosis of the wind payoff distribution indicates the presence of rare but very large payoff realizations. These results suggest that wind PPAs expose market participants to substantially greater uncertainty than their solar counterparts.

\begin{table}[H]
\centering
\caption{Tail risk measures for the payoff distribution of the fixed-price solar PPA evaluated at the fair strike price.}
\label{tab:pv_solar_risk}
\begin{tabular}{lrr}
\hline
Significance Level & T-VaR (€) & T-ES (€) \\
\hline
1\%   & -3,539,016.98 & -4,291,231.80 \\
2.5\% & -2,913,593.92 & -3,634,385.71 \\
5\%   & -2,376,082.64 & -3,126,688.26 \\
95\%  & 2,282,747.73  & 3,053,144.42 \\
97.5\%& 2,800,416.84  & 3,599,848.77 \\
99\%  & 3,507,836.81  & 4,314,166.24 \\
\hline
\end{tabular}
\end{table}

\begin{table}[H]
\centering
\caption{Summary statistics of the payoff distribution for the fixed-price solar PPA evaluated at the fair strike price.}
\label{tab:pv_solar_moments}
\begin{tabular}{lr}
\hline
Statistic & Value \\
\hline
Mean      & -57,321.36 \\
Std. Dev. & 1,426,982.52 \\
Skewness  & 0.0343 \\
Kurtosis  & 1.0057 \\
\hline
\end{tabular}
\end{table}

\begin{table}[H]
\centering
\caption{Tail risk measures for the payoff distribution of the fixed-price wind PPA evaluated at the fair strike price.}
\label{tab:pv_wind_risk}
\begin{tabular}{lrr}
\hline
Significance Level & T-VaR (€) & T-ES (€) \\
\hline
1\%   & -22,313,652.68 & -36,267,534.08 \\
2.5\% & -13,062,669.32 & -24,621,488.20 \\
5\%   & -7,732,548.00  & -17,298,001.66 \\
95\%  & 8,007,047.74   & 16,948,259.98 \\
97.5\%& 12,950,085.86  & 23,831,574.20 \\
99\%  & 21,764,941.86  & 35,581,783.89 \\
\hline
\end{tabular}
\end{table}

\begin{table}[H]
\centering
\caption{Summary statistics of the payoff distribution for the fixed-price wind PPA evaluated at the fair strike price.}
\label{tab:pv_wind_moments}
\begin{tabular}{lr}
\hline
Statistic & Value \\
\hline
Mean      & 187,728.36 \\
Std. Dev. & 7,191,608.57 \\
Skewness  & -0.3557 \\
Kurtosis  & 72.0118 \\
\hline
\end{tabular}
\end{table}

Among the contract structures considered, stepped PPAs offer the most favorable payoff profile from the offtaker's perspective and therefore the most negative profile for the producer. For PV, the stepped contract delivers a strictly positive mean payoff with relatively limited downside risk, exhibiting only moderate asymmetry and tail thickness. This effect is even more pronounced for wind: the stepped wind contract achieves the highest expected payoff of any structure. However, this increase in expected profitability comes at the cost of substantially greater payoff uncertainty. The upper-tail quantiles are extremely large, skewness is strongly positive, and kurtosis is very high.

\begin{table}[H]
\centering
\caption{Tail risk measures for the payoff distribution of the stepped solar PPA evaluated at the fair strike price.}
\label{tab:step_solar_risk}
\begin{tabular}{lrr}
\hline
Significance Level & T-VaR (€) & T-ES (€) \\
\hline
1\%   & 148,211.32  & -85,625.33 \\
2.5\% & 471,086.50  & 165,211.38 \\
5\%   & 730,834.56  & 389,951.18 \\
95\%  & 3,335,612.98 & 3,821,336.34 \\
97.5\%& 3,614,657.77 & 4,186,834.23 \\
99\%  & 4,126,838.15 & 4,741,293.04 \\
\hline
\end{tabular}
\end{table}

\begin{table}[H]
\centering
\caption{Summary statistics of the payoff distribution for the stepped solar PPA evaluated at the fair strike price.}
\label{tab:step_solar_moments}
\begin{tabular}{lr}
\hline
Statistic & Value \\
\hline
Mean      & 2,016,173.28 \\
Std. Dev. & 838,922.54 \\
Skewness  & 0.2869 \\
Kurtosis  & 0.6089 \\
\hline
\end{tabular}
\end{table}


\begin{table}[H]
\centering
\caption{Tail risk measures for the payoff distribution of the stepped wind PPA evaluated at the fair strike price.}
\label{tab:step_wind_risk}
\begin{tabular}{lrr}
\hline
Significance Level & T-VaR (€) & T-ES (€) \\
\hline
1\%   & -954,508.12   & -3,193,350.52 \\
2.5\% & -28,903.86    & -1,481,455.84 \\
5\%   & 77,537.41     & -724,097.69 \\
95\%  & 16,761,183.62 & 30,200,776.36 \\
97.5\%& 24,230,254.49 & 40,482,219.57 \\
99\%  & 37,130,427.51 & 56,550,940.87 \\
\hline
\end{tabular}
\end{table}

\begin{table}[H]
\centering
\caption{Summary statistics of the payoff distribution for the stepped wind PPA evaluated at the fair strike price.}
\label{tab:step_wind_moments}
\begin{tabular}{lr}
\hline
Statistic & Value \\
\hline
Mean      & 4,568,480.50 \\
Std. Dev. & 7,901,560.49 \\
Skewness  & 6.3838 \\
Kurtosis  & 82.2968 \\
\hline
\end{tabular}
\end{table}

The reverse-collar structures mitigate part of the downside exposure for both parties. This effect is particularly evident for the PV contract, whose lower-tail quantiles become less extreme and whose volatility decreases relative to the fixed-price specification. For wind power generation, the collar also reduces downside risk, although the payoff distribution remains highly asymmetric and leptokurtic. The positive skewness observed for both reverse-collar contracts indicates that the contractual design moves the payoff profile toward favorable outcomes, while the very high kurtosis of the wind contract highlights the presence of significant tail risk despite the protection offered by the collar mechanism.

Overall, solar PPAs deliver more stable and predictable payoffs, whereas wind PPAs offer larger positive potential at the cost of substantially higher volatility and tail risk. Across contract types, transitioning from fixed-price to reverse-collar and finally to stepped structures progressively skews the payoff distribution toward positive outcomes. Stepped contracts provide the greatest revenue enhancement for offtakers, but this benefit is accompanied by a larger exposure to extreme payoff realizations.

\begin{table}[htbp]
\centering
\caption{Tail risk measures for the payoff distribution of the reverse-collar solar PPA evaluated at the fair strike price.}
\label{tab:rc_solar_risk}
\begin{tabular}{lrr}
\hline
Significance Level & T-VaR (€) & T-ES (€) \\
\hline
1\%   & -1,998,227.60 & -2,617,071.41 \\
2.5\% & -1,475,504.55 & -2,061,039.80 \\
5\%   & -1,135,043.65 & -1,671,891.90 \\
95\%  & 2,216,492.61  & 2,778,229.67 \\
97.5\%& 2,660,734.71  & 3,141,017.10 \\
99\%  & 3,192,456.02  & 3,517,731.31 \\
\hline
\end{tabular}
\end{table}

\newpage

\begin{table}[htbp]
\centering
\caption{Summary statistics of the payoff distribution for the reverse-collar solar PPA evaluated at the fair strike price.}
\label{tab:rc_solar_moments}
\begin{tabular}{lr}
\hline
Statistic & Value \\
\hline
Mean      & 228,572.21 \\
Std. Dev. & 1,061,993.12 \\
Skewness  & 0.5297 \\
Kurtosis  & 0.6860 \\
\hline
\end{tabular}
\end{table}

\begin{table}[htbp]
\centering
\caption{Tail risk measures for the payoff distribution of the reverse-collar wind PPA evaluated at the fair strike price.}
\label{tab:rc_wind_risk}
\begin{tabular}{lrr}
\hline
Significance Level & T-VaR (€) & T-ES (€) \\
\hline
1\%   & -12,912,425.30 & -22,124,492.01 \\
2.5\% & -7,753,338.49  & -14,665,220.09 \\
5\%   & -4,712,623.24  & -10,297,291.21 \\
95\%  & 7,536,643.52   & 16,084,045.16 \\
97.5\%& 12,040,107.81  & 22,570,580.80 \\
99\%  & 20,251,622.73  & 33,249,590.42 \\
\hline
\end{tabular}
\end{table}

\begin{table}[htbp]
\centering
\caption{Summary statistics of the payoff distribution for the reverse-collar wind PPA evaluated at the fair strike price.}
\label{tab:rc_wind_moments}
\begin{tabular}{lr}
\hline
Statistic & Value \\
\hline
Mean      & 631,254.90 \\
Std. Dev. & 5,703,299.48 \\
Skewness  & 1,4579 \\
Kurtosis  & 100.4168 \\
\hline
\end{tabular}
\end{table}

\begin{figure}[H]
\centering

\begin{subfigure}[b]{0.48\textwidth}
    \centering
    \includegraphics[width=\textwidth]{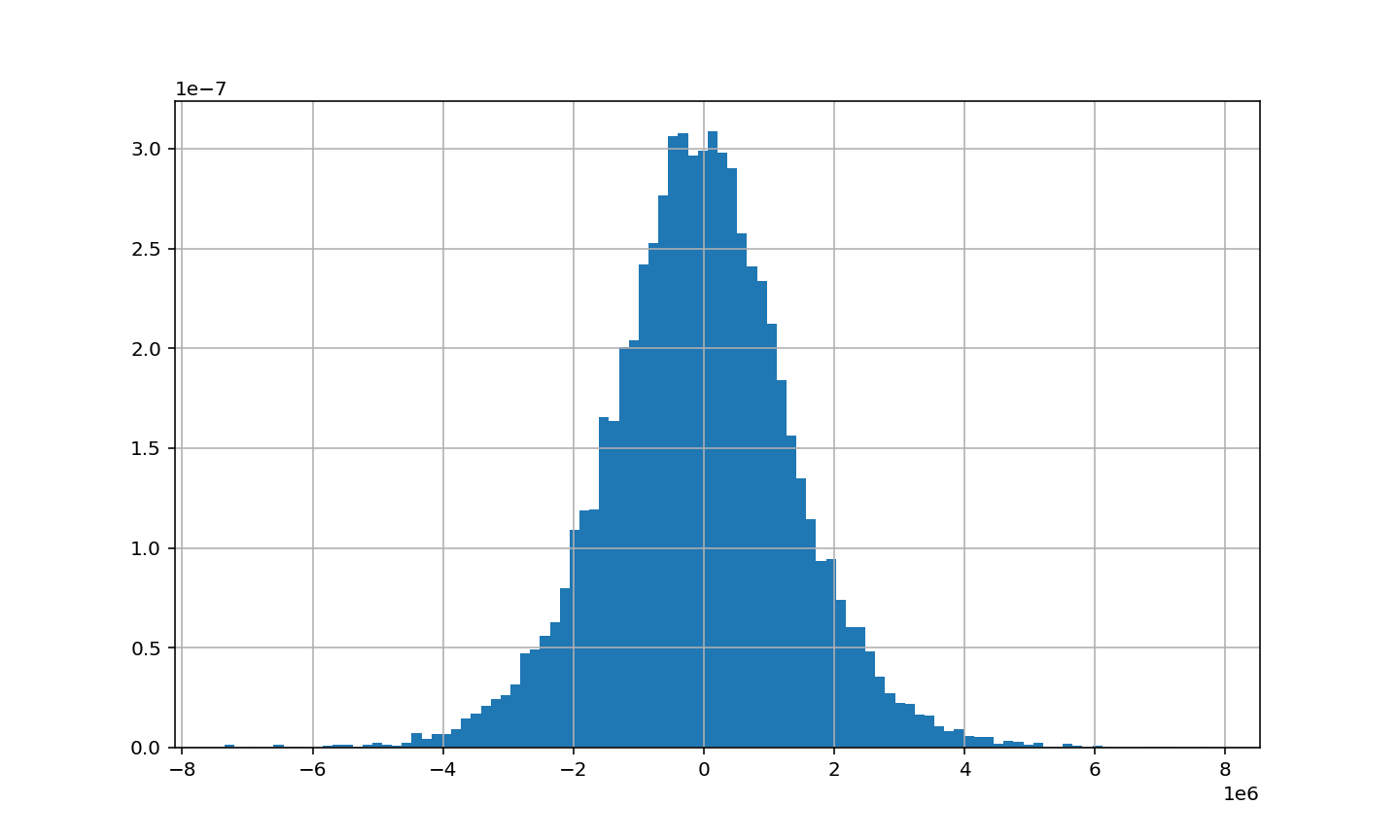}
    \caption{Fixed-price solar PPA.}
    \label{fig:ppa_plain_vanilla_solar_payoff}
\end{subfigure}
\hfill
\begin{subfigure}[b]{0.48\textwidth}
    \centering
    \includegraphics[width=\textwidth]{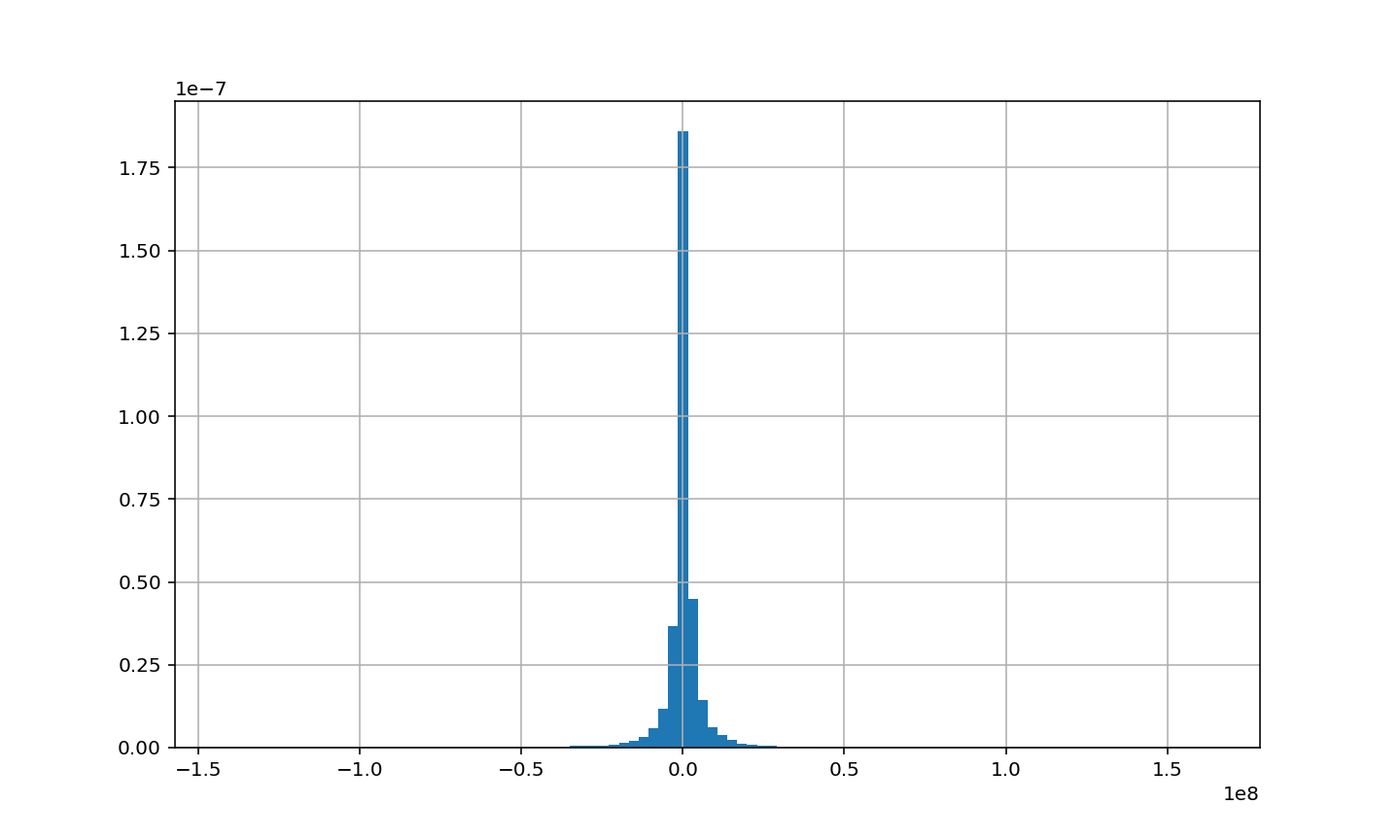}
    \caption{Fixed-price wind PPA.}
    \label{fig:ppa_plain_vanilla_wind_payoff}
\end{subfigure}

\vspace{0.5cm}

\begin{subfigure}[b]{0.48\textwidth}
    \centering
    \includegraphics[width=\textwidth]{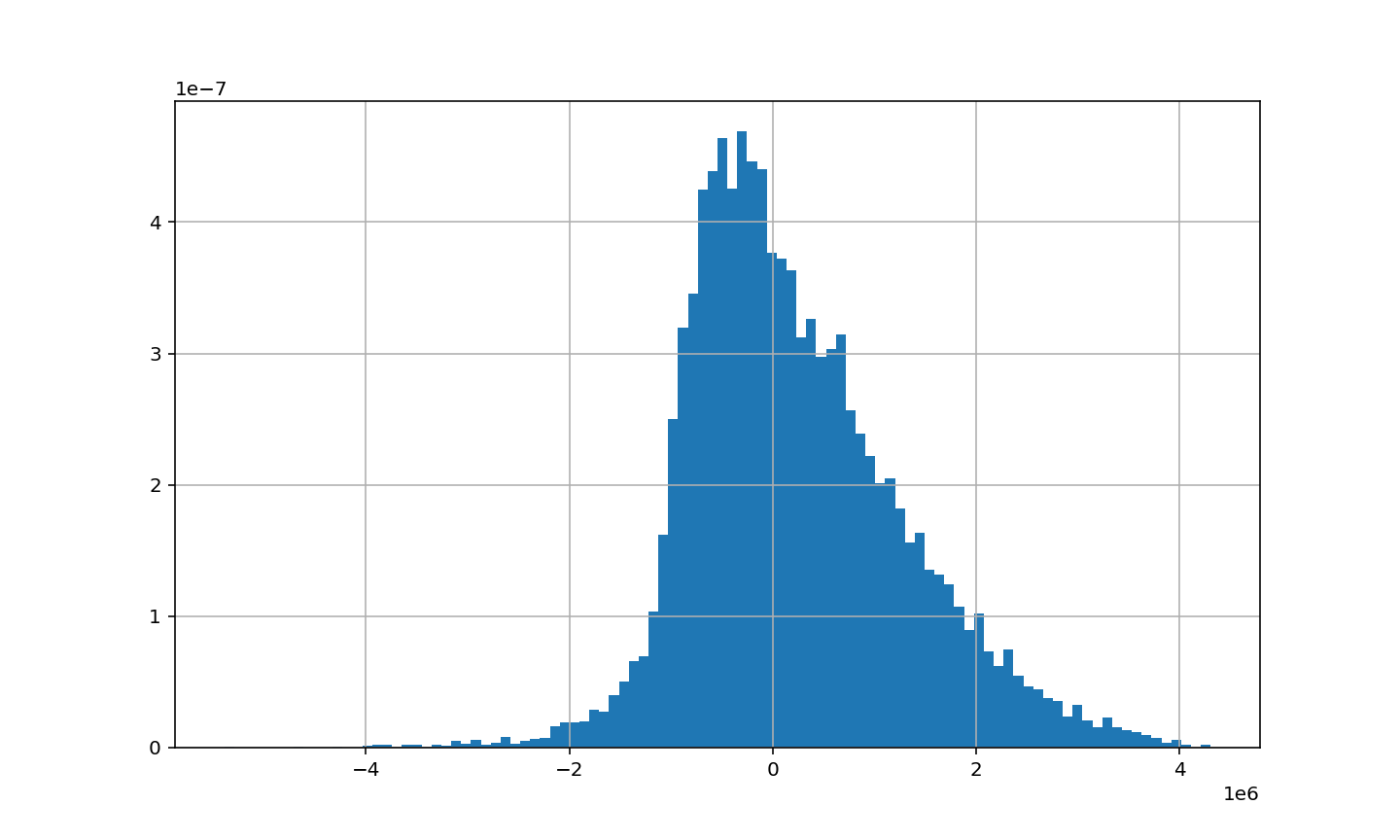}
    \caption{Reverse-collar solar PPA.}
    \label{fig:ppa_reverse_collar_solar_payoff}
\end{subfigure}
\hfill
\begin{subfigure}[b]{0.48\textwidth}
    \centering
    \includegraphics[width=\textwidth]{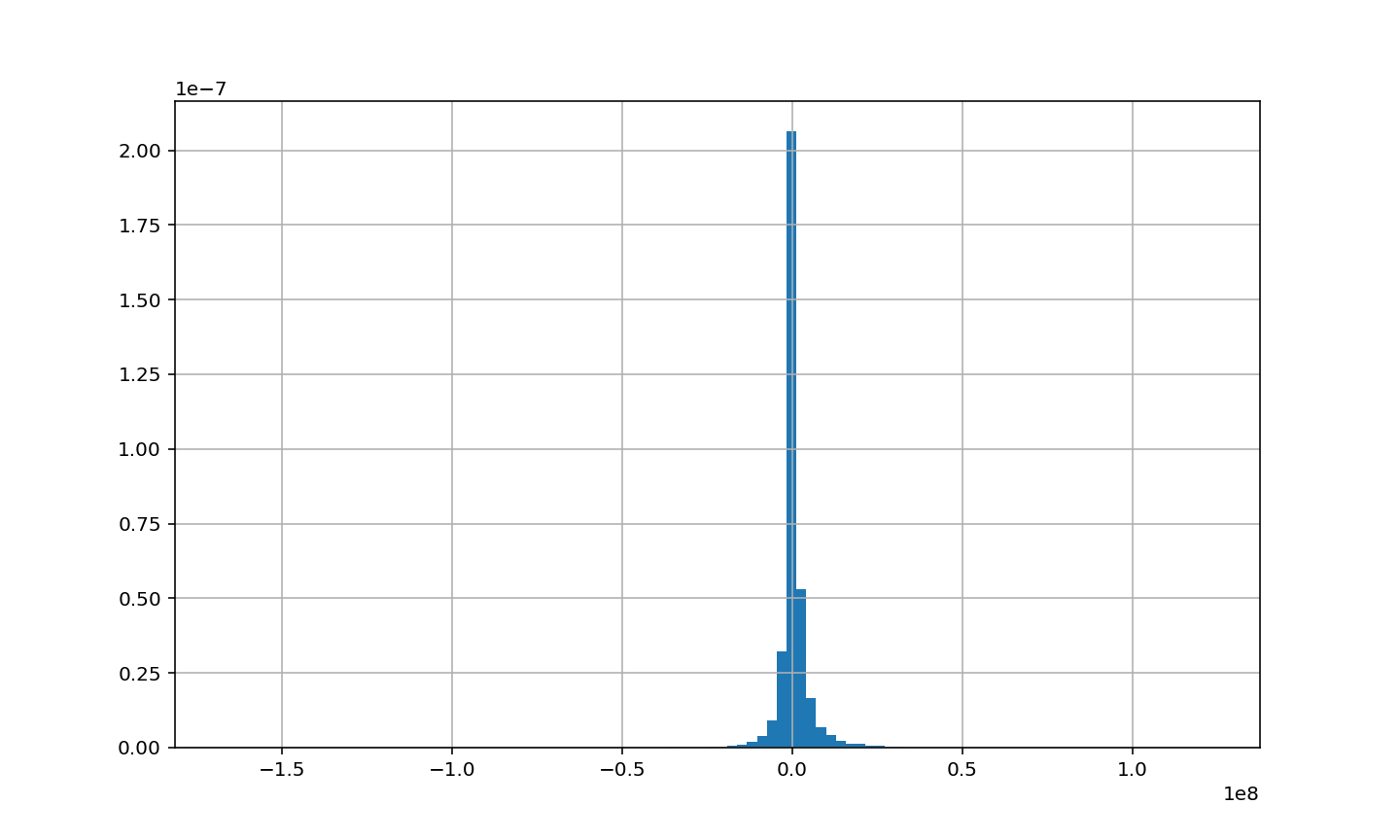}
    \caption{Reverse-collar wind PPA.}
    \label{fig:ppa_reverse_collar_wind_payoff}
\end{subfigure}

\vspace{0.5cm}

\begin{subfigure}[b]{0.48\textwidth}
    \centering
    \includegraphics[width=\textwidth]{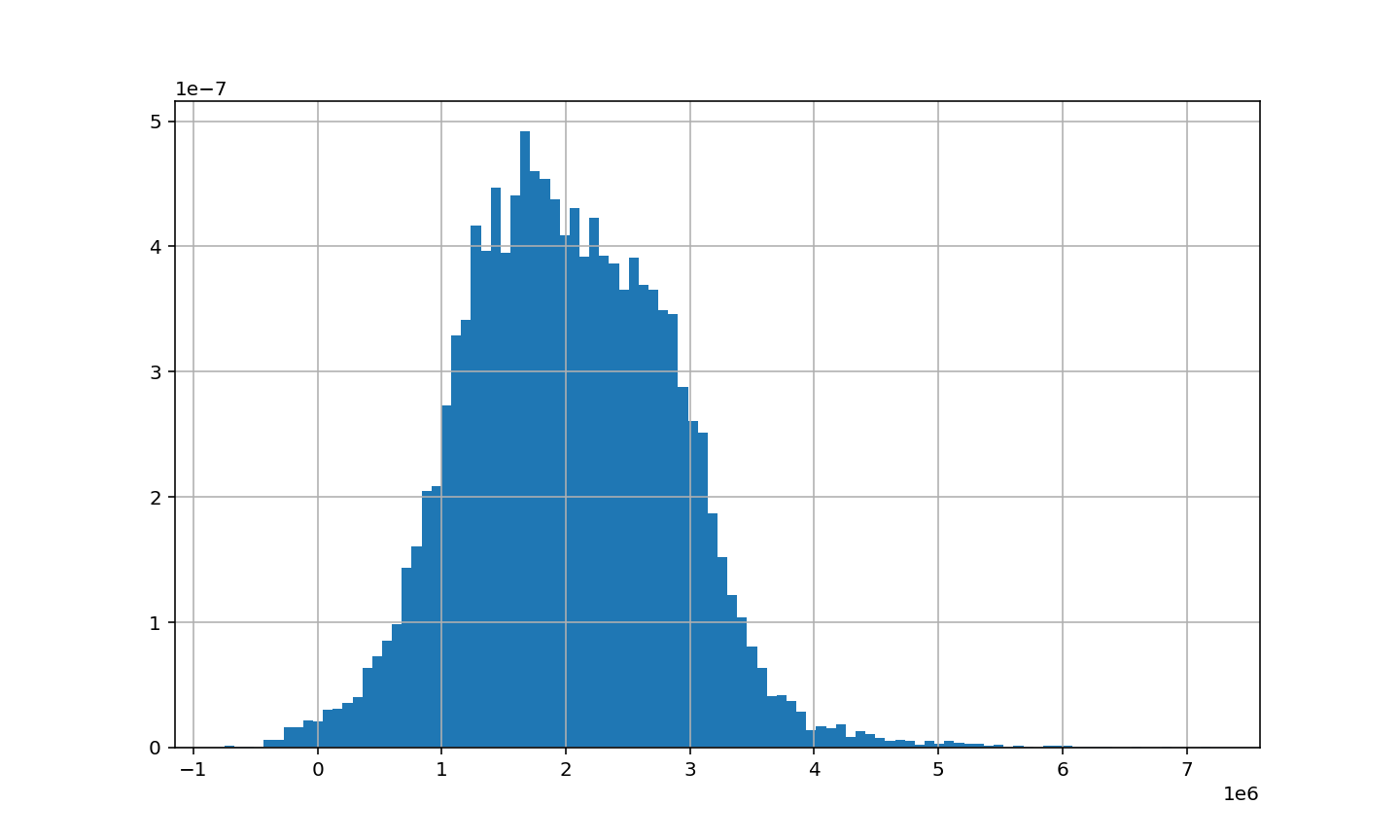}
    \caption{Stepped solar PPA.}
    \label{fig:stepped_solar_payoff}
\end{subfigure}
\hfill
\begin{subfigure}[b]{0.48\textwidth}
    \centering
    \includegraphics[width=\textwidth]{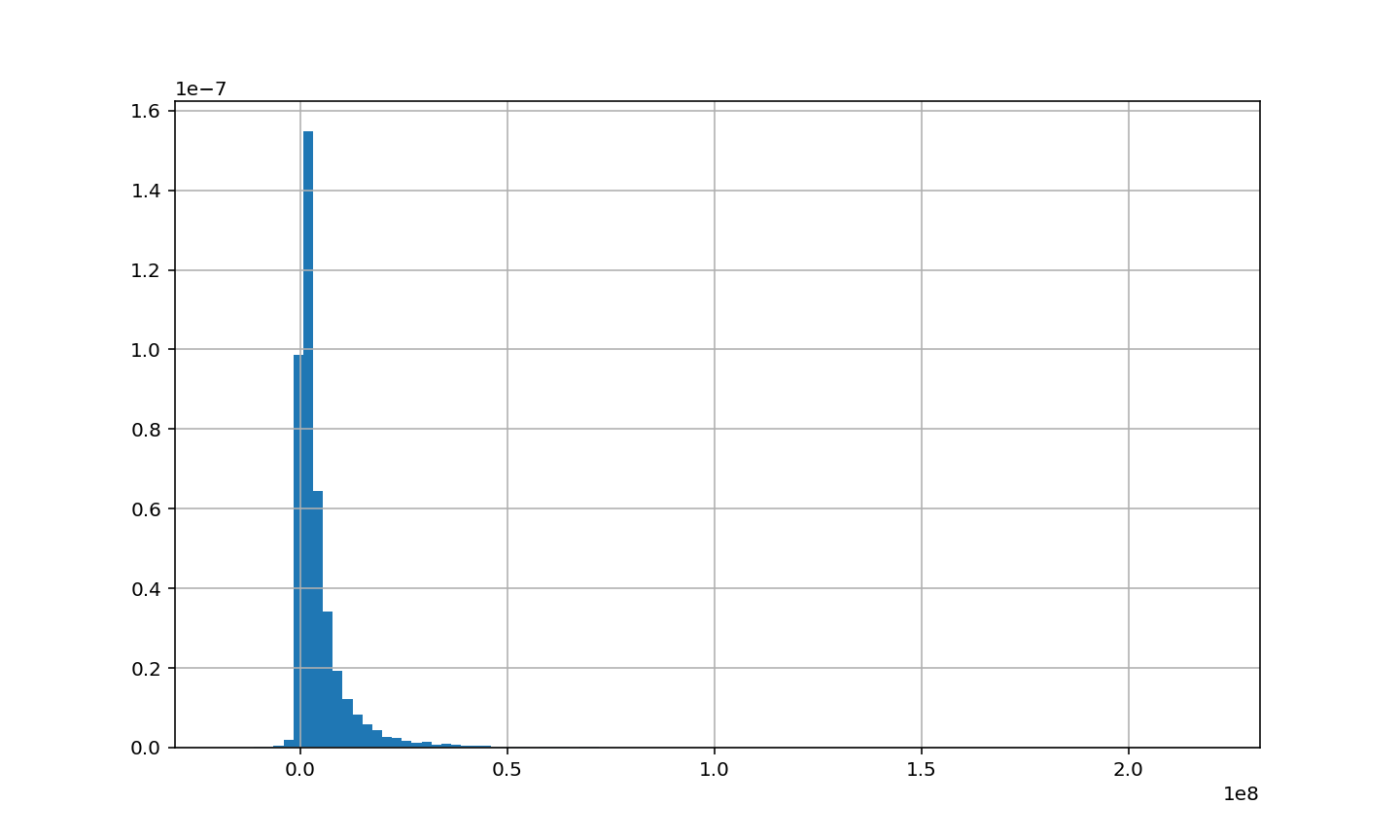}
    \caption{Stepped wind PPA.}
    \label{fig:stepped_wind_payoff}
\end{subfigure}

\caption{Simulated terminal payoff distributions of the fair-value PPA contracts for PV and wind. The first row holds the fixed-price PPA, the second row the reverse-collar PPA, and the third row the stepped PPA.}
\label{fig:ppa_payoff_distributions}
\end{figure}

\section{Conclusions}
\label{sec:conclusions}
The proposed main contributions of this work were four: first, formalizing, in financial mathematics terms, the payoff structure of the fixed-price renewable PPA types currently in existence. Second, to propose a pricing approach, for those additional contractual specifications, which relies on the concept of financial fairness, in order to provide the regulator with a tool to detect and quantify deviations from market fairness, so that proper action can be taken to understand the reasons behind them or to correct potential inefficiencies. Third, we attempted to address the research gap highlighted in \cite{zhao2026hybrid}, by developing a risk-assessment methodology able to balance financial soundness with the complexity of the PPA and of its underlying risk factors. Finally, we introduced a new methodology to model the daily solar irradiance for which we could limit the number of parameters and complexity, making it useful for daily simulations and practical applications, while being able to represent the main features of the process. 

The empirical analysis, with simulations based on the Italian electricity market, showed that the fair prices of PV PPAs are lower than those for wind. Additionally, the analysis of the terminal contract payoffs  highlights significant differences in the risk-return profiles of the considered PPA structures. Among the contractual designs, reverse-collar PPAs provide a partial reduction in downside exposure for the offtaker, while stepped PPAs generate the highest expected payoffs but also the greatest payoff asymmetry and exposure to extreme outcomes. Overall, the choice of contract structure involves a trade-off between risk mitigation and participation in favorable market conditions. Additionally, solar-based contracts generally exhibit lower volatility and more stable payoff distributions, while wind-based contracts offer greater upside potential at the cost of substantially higher tail risk. 

There are however a few limitations to our work. The first is the absence of a comparison with the prices of real PPAs, which was not possible due to the lack of sufficient data. The second is the preliminary nature of our risk assessment approach, which does not yet address the counterparty credit risk of the producer and the offtaker. This is an essential risk factor for over-the-counter contracts. We reserve this exploration for future research. Further research could also be aimed at extending this framework to additional contractual specifications, offering an assessment of their different risk profiles.

\newpage

\bibliographystyle{plainnat}
\bibliography{references}

\section{Appendix A}
\subsection{Preliminary analysis}
This section of the appendix holds a preliminary analysis of the time series of the underlying risk factors, highlighting the features that are captured by the selected continuous-time models.

\subsubsection{Electricity price preliminary analysis}
\label{sec:Electricity_price_preliminary_analysis}
\begin{figure}[H]
    \centering
    \includegraphics[width=0.75\linewidth]{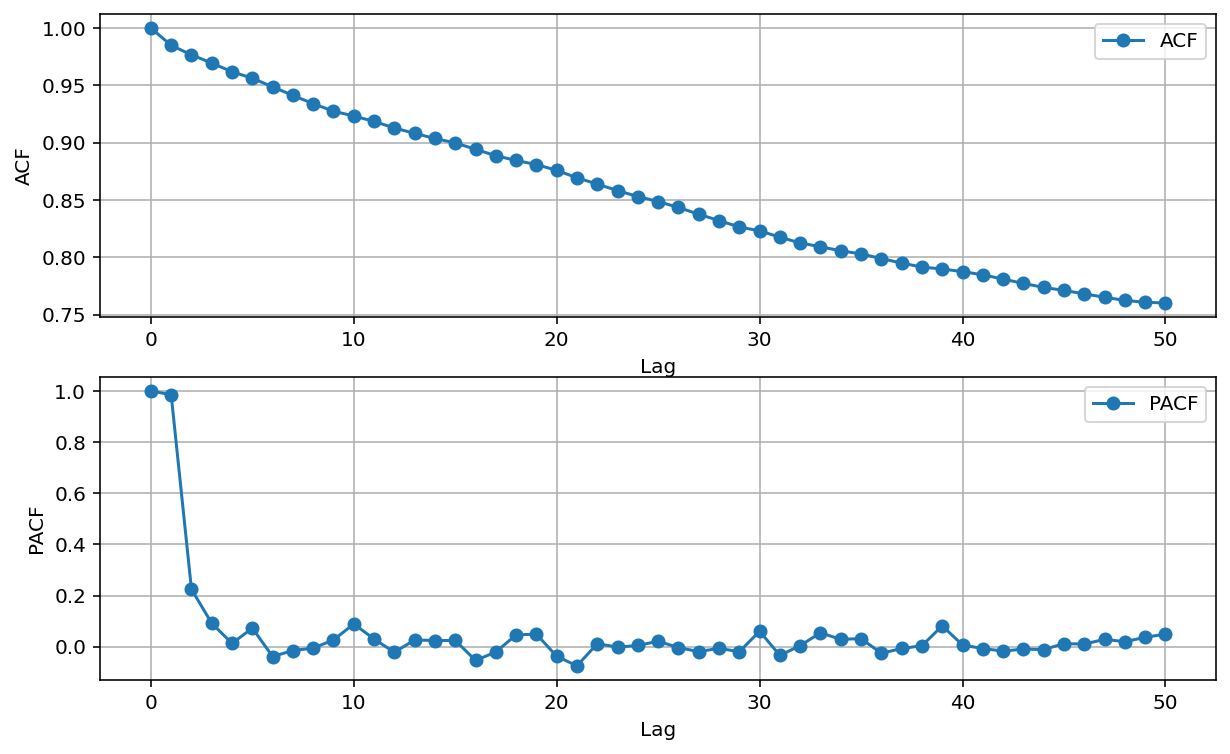}
    \caption{Autocorretion and Partial Autocorrelation of $X_t$}
    \label{fig:pun_acf_pacf}
\end{figure}

\begin{figure}[H]
    \centering
    \includegraphics[width=0.75\linewidth]{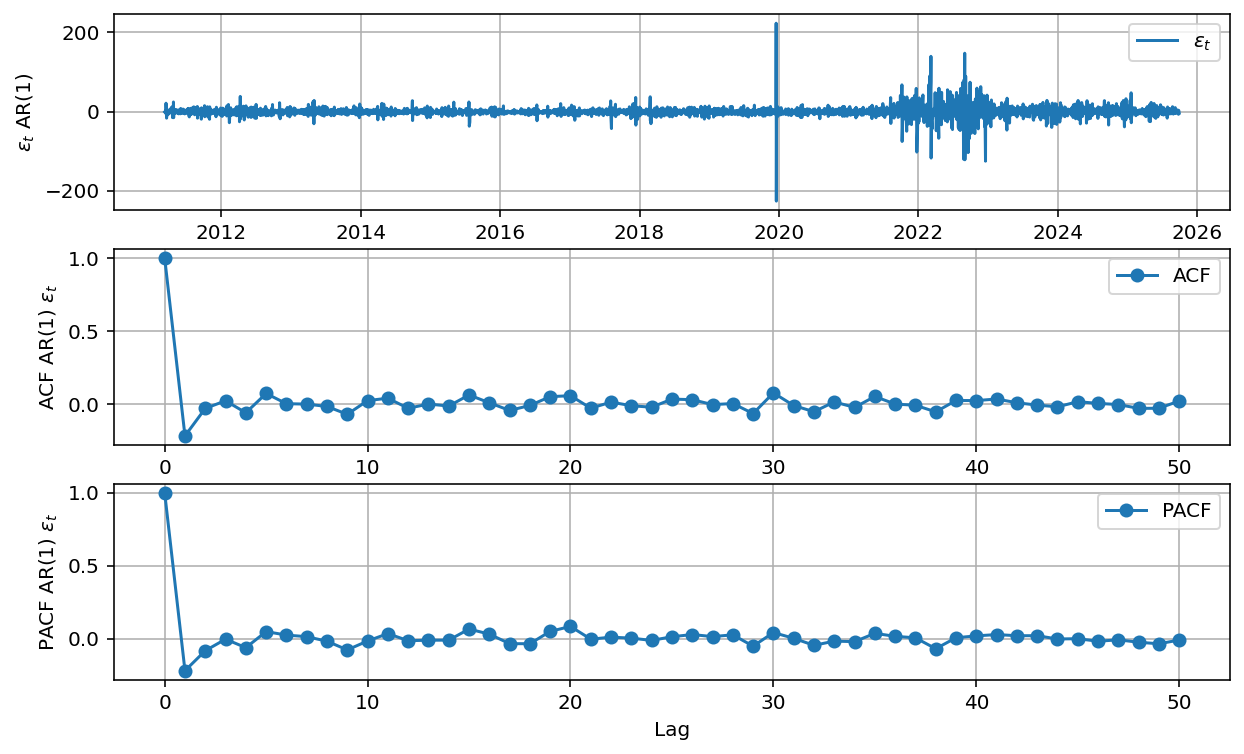}
    \caption{ACF and PACF of AR(1) $X_t$ residuals ($\epsilon_t$)}
    \label{fig:ar1_eps_pun_acf_pacf}
\end{figure}

In this section we highlight the features of data which lead to the choice of the model in Eq. \ref{eq:non_seasonal_part_electricity_price}. Such model, when discretized, is auto-regressive (AR) of order 1, with stochastic volatility. As we can see from Figure \ref{fig:pun_acf_pacf} the non-seasonal component of the electricity price, $X_t$, shows serial dependence with the past, leading us to the selection of an auto-regressive model. Once an AR(1) model is fit to the data, the auto-correlation substantially drops, as visible in Figure \ref{fig:ar1_eps_pun_acf_pacf}. 

\newpage

\begin{figure}[H]
    \centering
    \includegraphics[width=0.75\linewidth]{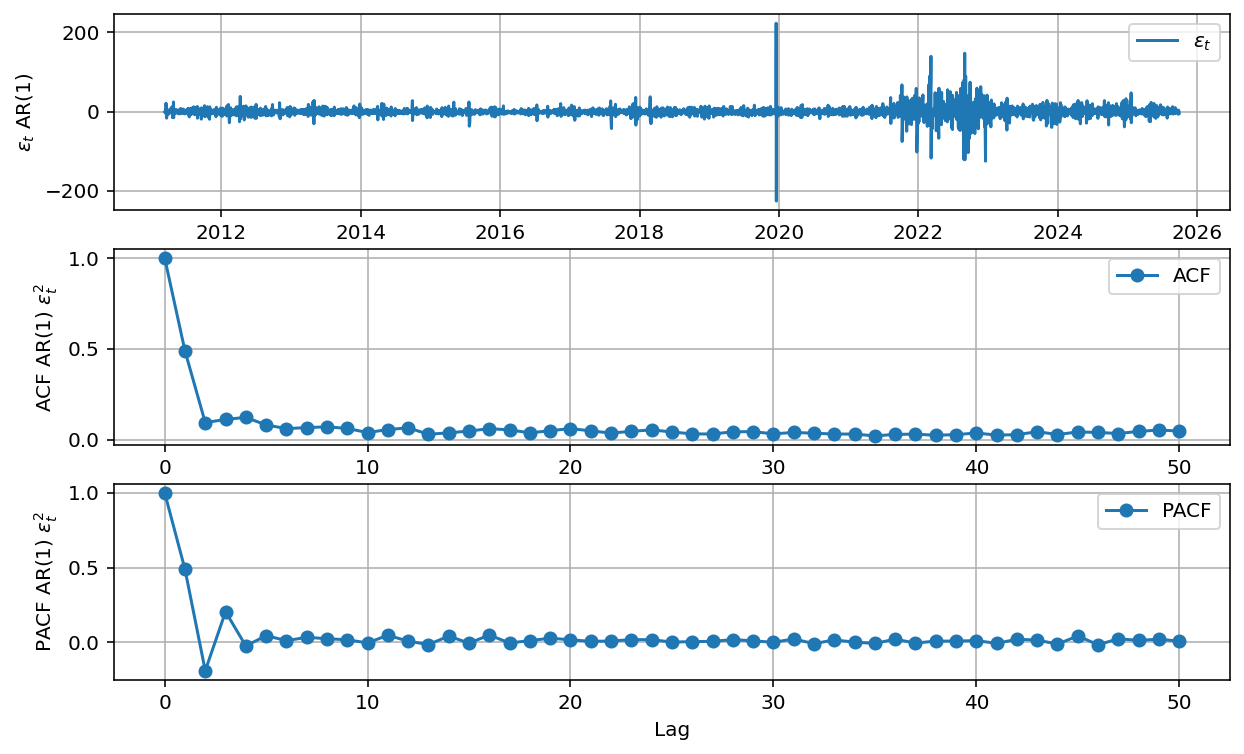}
    \caption{ACF and PACF of AR(1) $X_t$ squared residuals ($\epsilon_t^2$)}
    \label{fig:ar1_eps_square_pun_acf_pacf}
\end{figure}

\begin{figure}[H]
    \centering
    \includegraphics[width=0.75\linewidth]{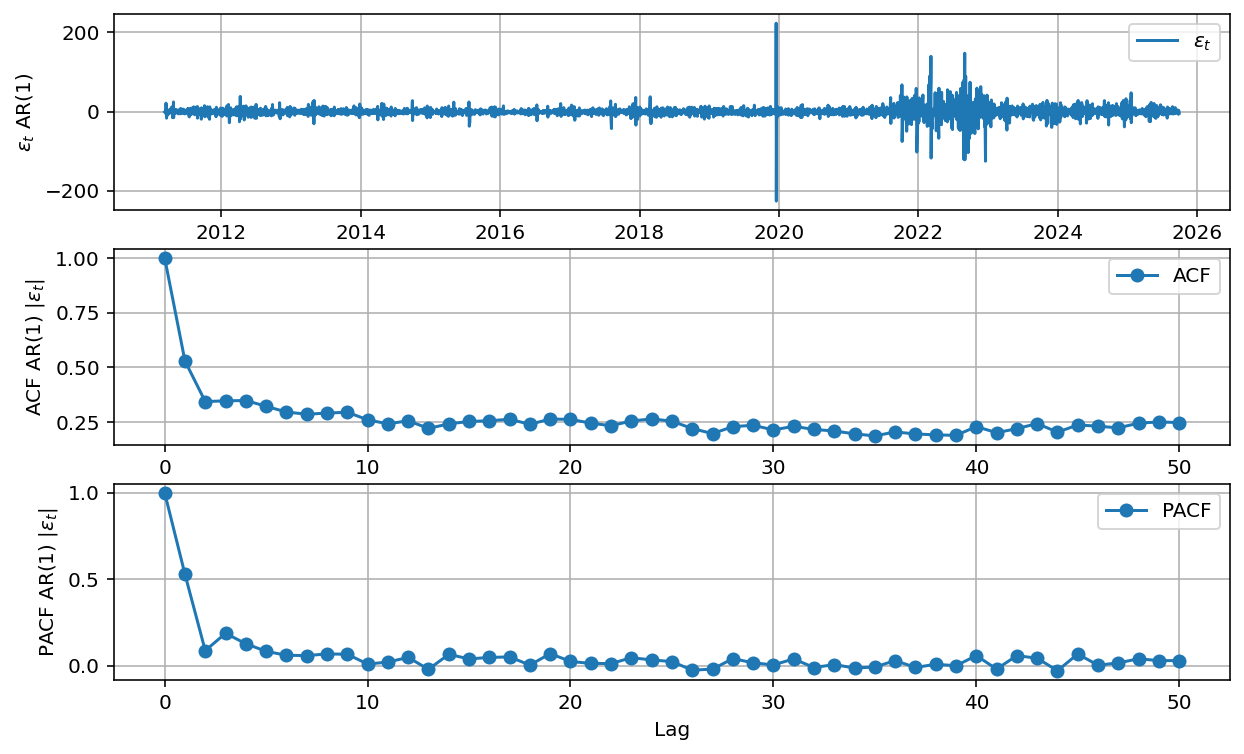}
    \caption{ACF and PACF of the absolute value of AR(1) $X_t$ residuals ($|\epsilon_t|$)}
    \label{fig:ar1_abs_eps_pun_acf_pacf}
\end{figure}

Next, to check for the presence of stochastic volatility, we show that the AR(1) residuals are not independent with the past even though they are uncorrelated.
Figures \ref{fig:ar1_eps_square_pun_acf_pacf} and \ref{fig:ar1_abs_eps_pun_acf_pacf} show that applying continuous (and therefore measurable) transformations to the residuals, such as the square and the absolute value, does not eliminate serial dependence. The persistence observed in these transformed series is indicative of stochastic volatility. Overall, these analyses motivate the adoption of models that incorporate both mean-reverting dynamics and stochastic volatility. We acknowledge that an SDE-based framework may not be sufficiently rich to capture all the features of the underlying process. Nevertheless, more sophisticated specifications would significantly increase the computational burden and complicate simulation procedures, particularly when modeling the dependence structure with other variables for the purpose of pricing financial contracts. In this work, we attempt to balance model richness with the complexity of the task that needs to be performed.



\subsubsection{Wind speed preliminary analysis}
\label{sec:Wind_preliminary_analysis}
\begin{figure}[H]
    \centering
    \includegraphics[width=0.75\linewidth]{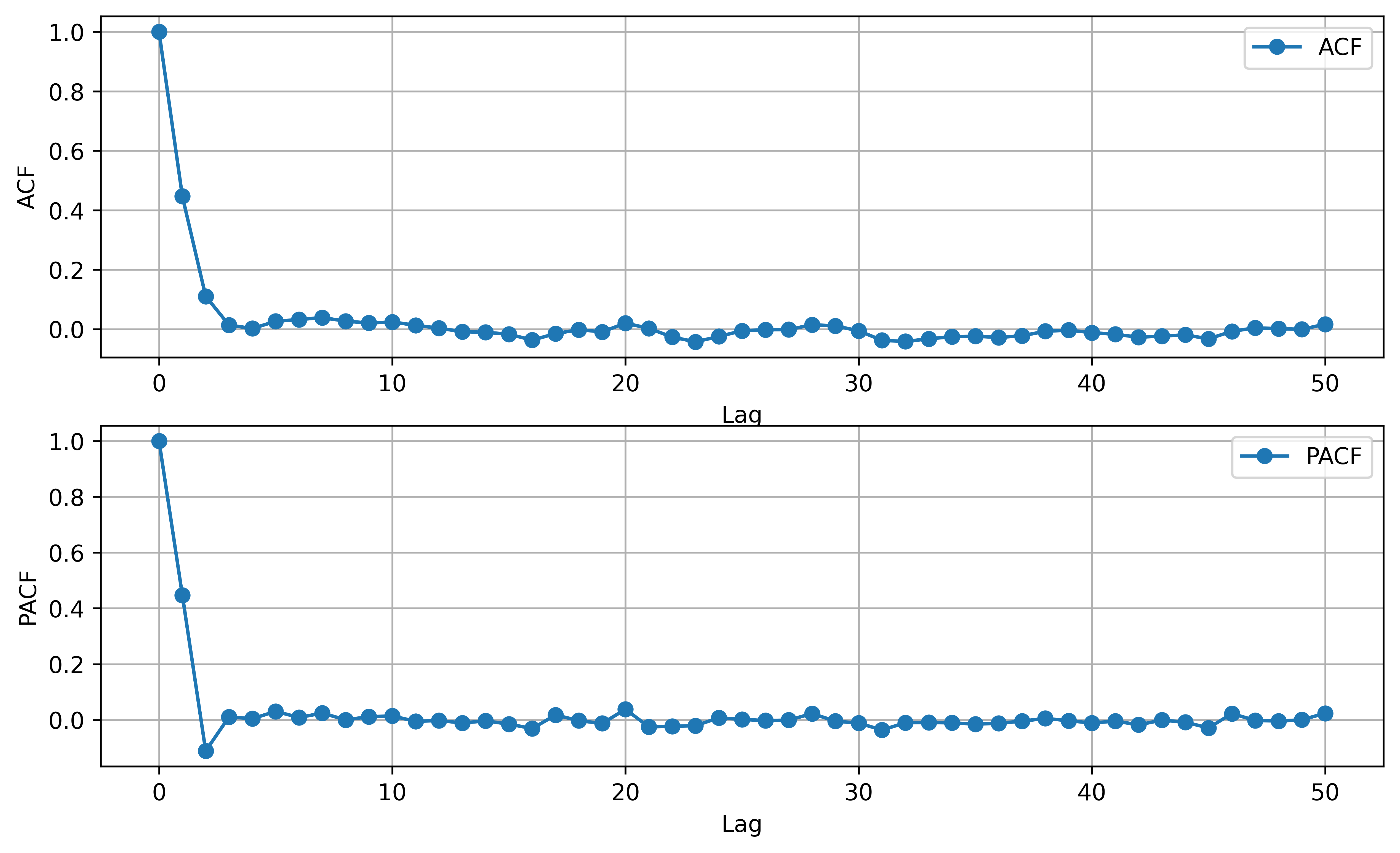}
    \caption{Autocorretion and Partial Autocorrelation of $Y_t$}
    \label{fig:wind_acf_pacf}
\end{figure}

\begin{figure}[H]
    \centering
    \includegraphics[width=0.75\linewidth]{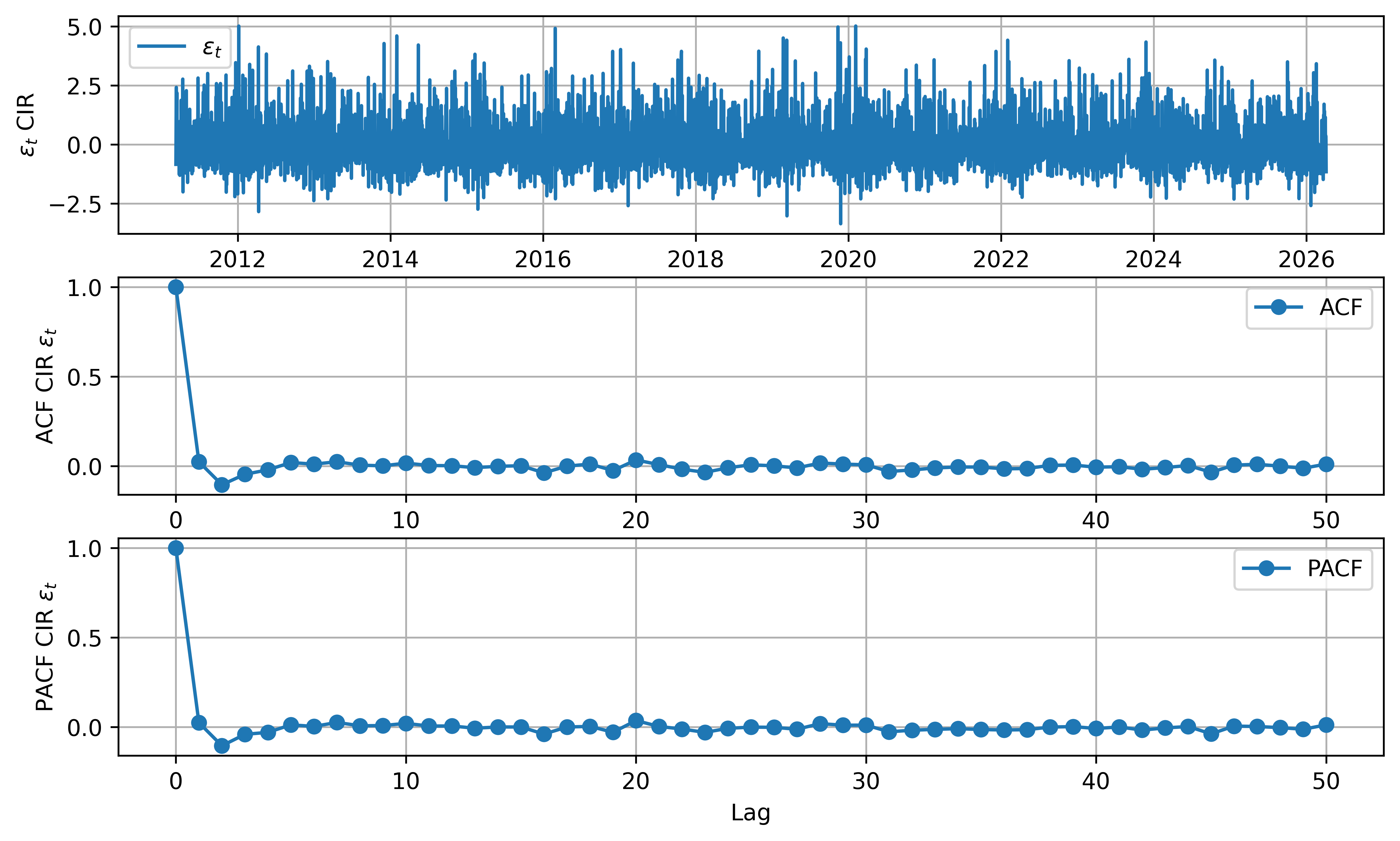}
    \caption{ACF and PACF of CIR residuals ($\epsilon_t$) of the non-seasonal component of average daily wind speed}
    \label{fig:cir_wind_acf_pacf}
\end{figure}

In this section we highlight the features of the data which lead to the selection of the model in Eq. \ref{eq:non_seasonal_part_wind} for the non-seasonal component of average daily wind speed, $Y_t$. The main features of this model are positivity and mean-reversion. The positivity of wind speed values can be immediately confirmed by observing the time series in Figure \ref{fig:wind_speed_data_plot}. As for the mean-reversion property, we make use of the discretization of the C.I.R. process, which satisfies 
\begin{equation*}
    \frac{Y_{t+\Delta t} - Y_t}{\sqrt{Y_t}} = \frac{\kappa \bar{Y}\Delta t}{\sqrt{Y_t}} - \kappa \sqrt{Y_t} \Delta t + \sigma \sqrt{\Delta t} \epsilon_t,
\end{equation*}
where $\epsilon_t$ is the i.i.d. noise component.
As we can see from Figure \ref{fig:wind_acf_pacf}, the non-seasonal component of average daily wind speed exhibits serial dependence with the past, but once the CIR model is fit to the data, the auto-correlation of the standardized residuals substantially drops, as visible in Figure \ref{fig:cir_wind_acf_pacf}. 

\subsubsection{GHI preliminary analysis}
\label{sec:ghi_preliminary_analysis}

The discretized version of the model for $G_t$ in Eq. \ref{eq:ou_part_solar} is an AR(1) process with stochastic volatility. In this Section we illustrate why this proposal agrees with the features of the time series at hand: we show that the residual of the AR(1) model fit to the data are not independent with the past, even though they are uncorrelated. The persistence of serial dependence in these transformed series is indicative of stochastic volatility.

\begin{figure}[H]
    \centering
    \includegraphics[width=0.75\linewidth]{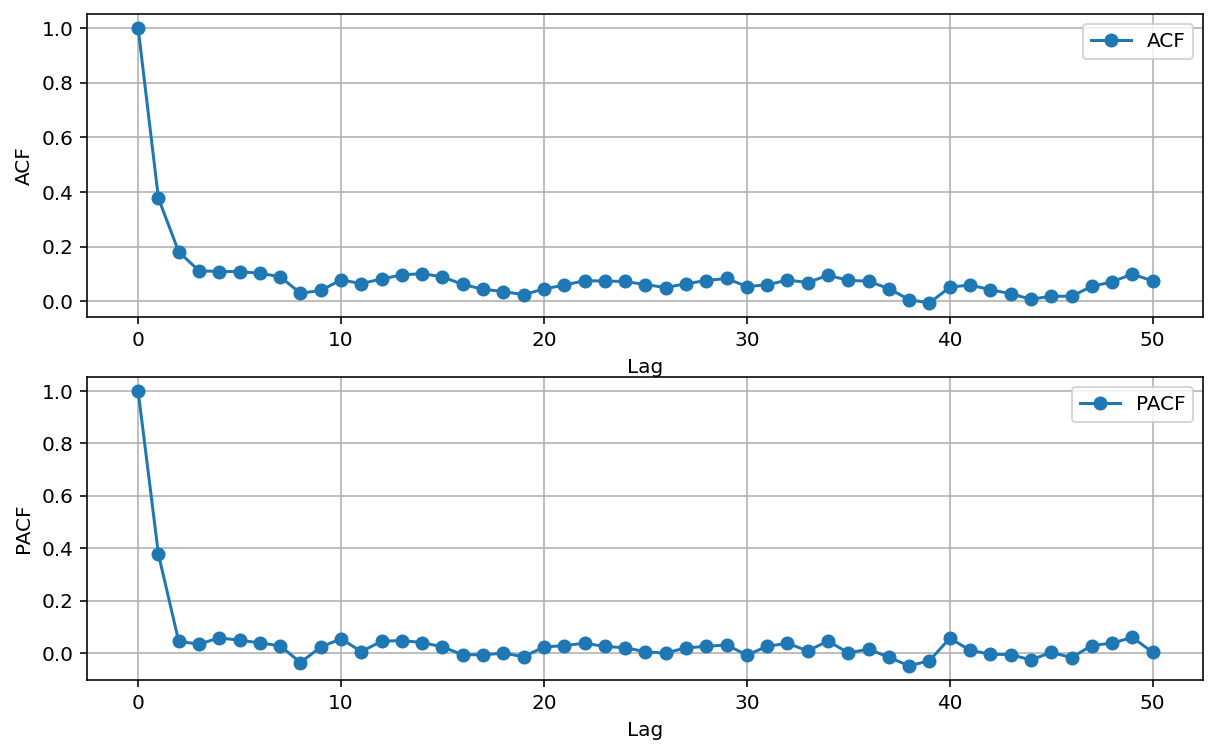}
    \caption{Autocorretion and Partial Autocorrelation of $G_t$}
    \label{fig:y_ghi_acf_pacf}
\end{figure}

\begin{figure}[H]
    \centering
    \includegraphics[width=0.75\linewidth]{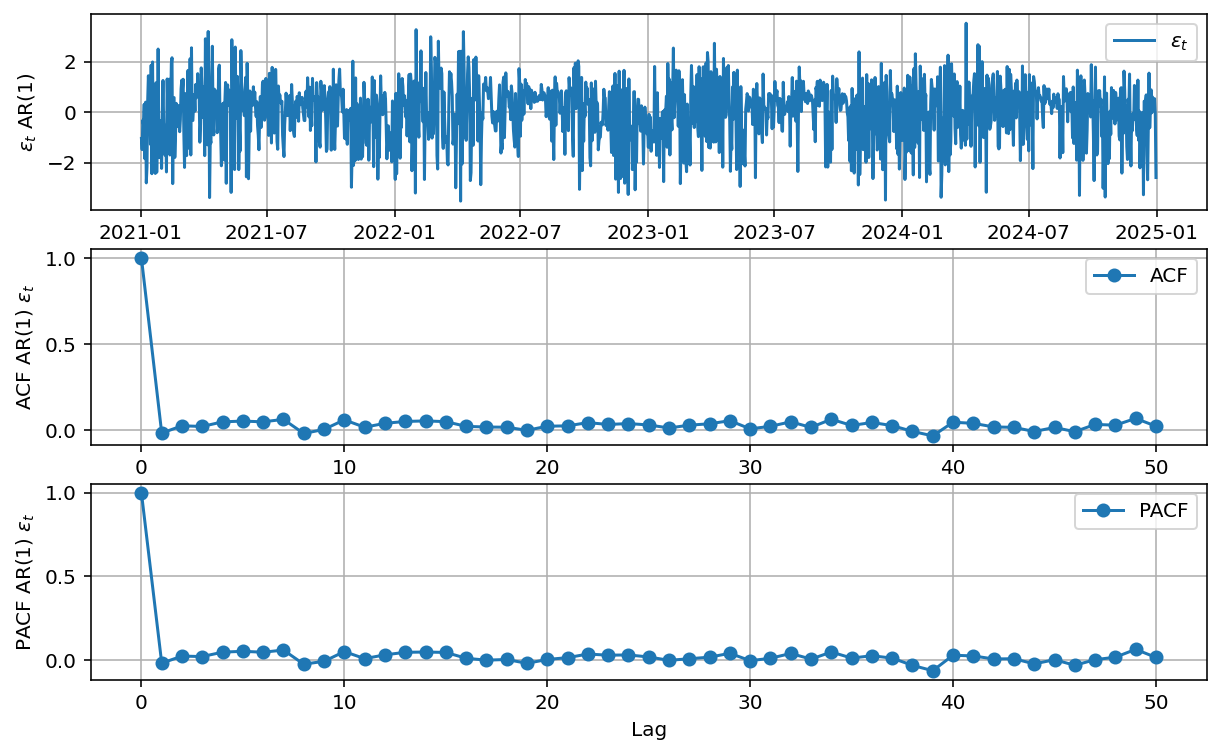}
    \caption{ACF and PACF of AR(1) $G_t$ residuals ($\epsilon_t$)}
    \label{fig:ar1_eps_y_ghi_acf_pacf}
\end{figure}

\begin{figure}[H]
    \centering
    \includegraphics[width=0.75\linewidth]{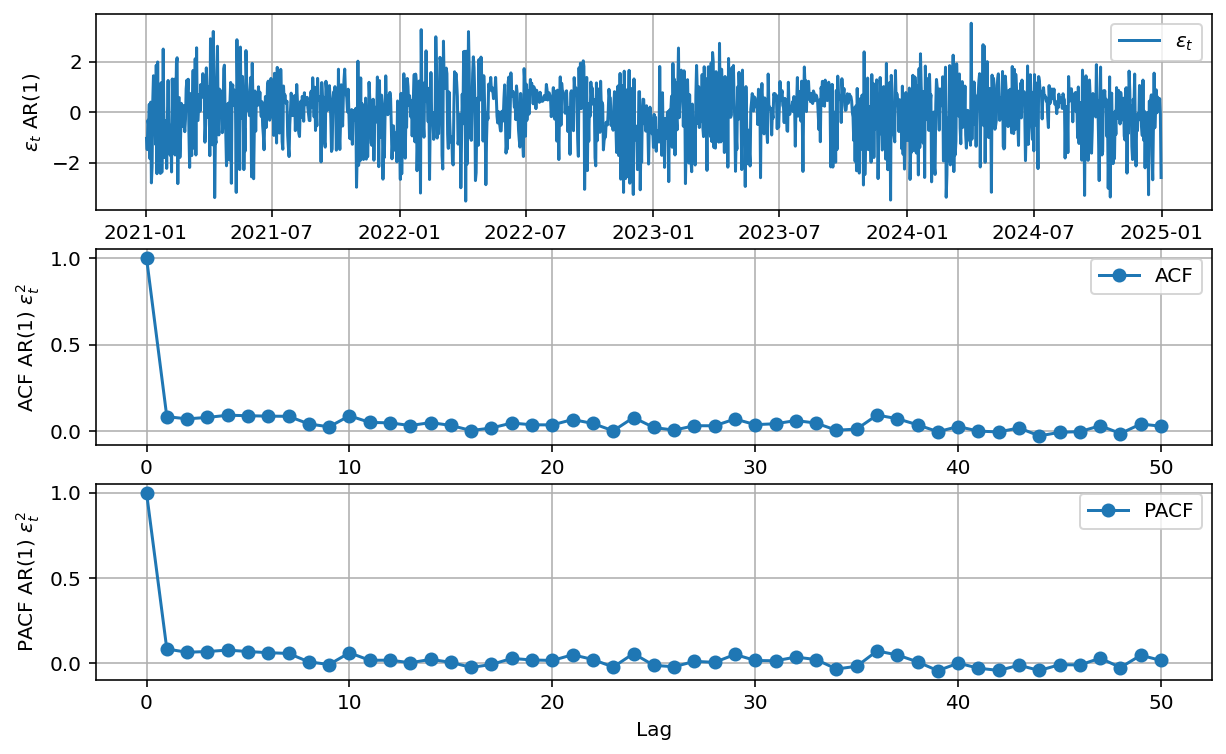}
    \caption{ACF and PACF of AR(1) $G_t$ squared residuals ($\epsilon_t^2$)}
    \label{fig:ar1_eps_square_y_ghi_acf_pacf}
\end{figure}

\begin{figure}[H]
    \centering
    \includegraphics[width=0.75\linewidth]{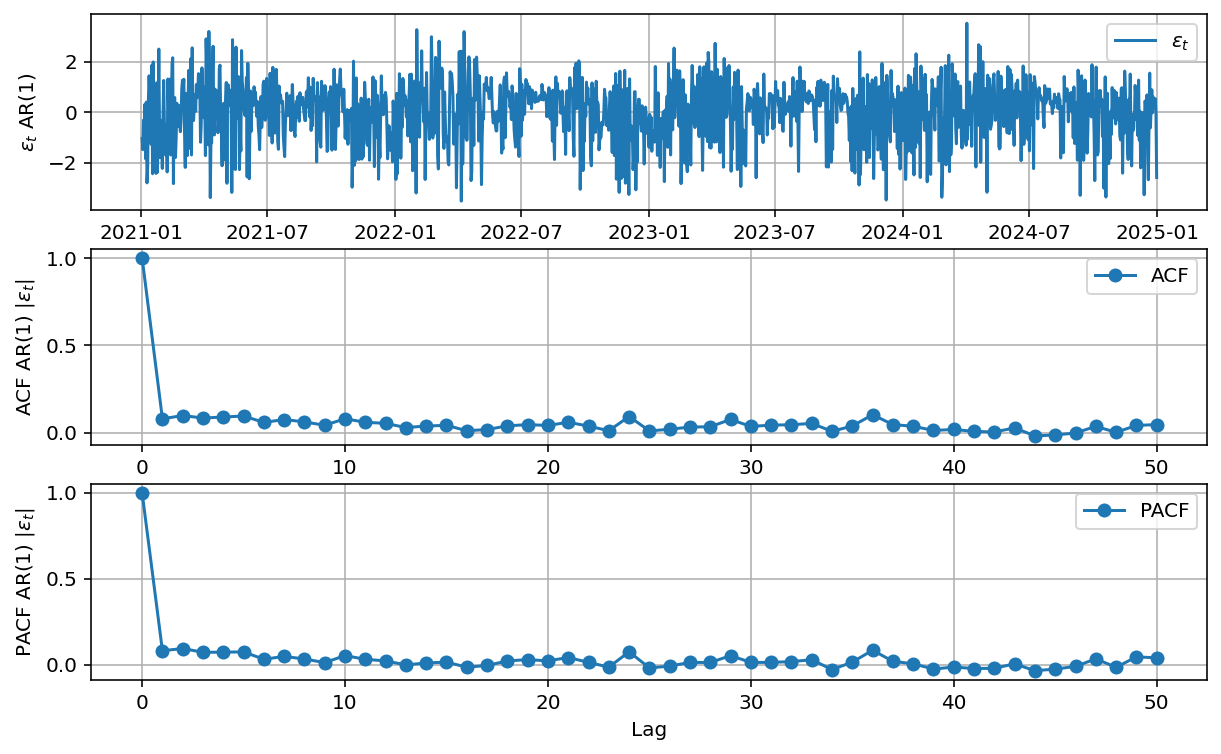}
    \caption{ACF and PACF of the absolute value of AR(1) $G_t$ residuals ($|\epsilon_t|$)}
    \label{fig:ar1_abs_eps_y_ghi_acf_pacf}
\end{figure}

As evidenced by Figure \ref{fig:y_ghi_acf_pacf}, $G_t$ shows serial dependence with the past. Once an AR(1) model is fit to the data, the auto-correlation substantially drops, as visible in Figure \ref{fig:ar1_eps_y_ghi_acf_pacf}. The residuals are uncorrelated but not independent, as shown in Figures \ref{fig:ar1_eps_square_y_ghi_acf_pacf} and \ref{fig:ar1_abs_eps_y_ghi_acf_pacf}.

\newpage

In fact, Figures \ref{fig:ar1_eps_square_y_ghi_acf_pacf} and \ref{fig:ar1_abs_eps_y_ghi_acf_pacf} show that by applying continuous transformations, such as the square and the absolute value, we still observe dependence with the past, which is why a model with mean reverting features and stochastic volatility is selected.

\newpage
\subsection{Terminal VaR and ES procedure}\label{app:algorithm_var_es}
Algorithm \ref{alg:mc_var_es} holds a step-by-step summary of the Terminal VaR and Terminal ES computation procedure.

\begin{algorithm}[H]
\caption{Terminal VaR and Expected Shortfall of PPA}
\label{alg:mc_var_es}
\begin{algorithmic}[1]
\State Input:
    \Statex \quad Settlement periods $i = 1,\dots,n$, each with $j = 1,\dots,m_i$ delivery sub-periods
    \Statex \quad Technology $\mathcal{T} \in \{\text{wind}, \text{PV}\}$
    \Statex \quad Contract type $\mathcal{C} \in \{\text{fixed-price}, \text{stepped}, \text{reverse collar}\}$
    \Statex \quad Strike price(s): $K$ (fixed-price), $\{K_l\}_{l=1}^L$ (stepped), or $(K,K_{min}, K_{max})$ (reverse collar)
    \Statex \quad Number of simulation paths $N$, Significance Level $\alpha = 0.95$
\State Output: $\widehat{\text{VaR}}_{0.95}$, $\widehat{\text{ES}}_{0.95}$ (positive = expected loss)
\Statex
\State Load technology specification
\If{$\mathcal{T} = \text{wind}$}
    \State Load joint model for $(S, W)$ calibrated to wind data
    \State Define production function $Q = Q(T, W)$ and set $P = W$
\ElsIf{$\mathcal{T} = \text{PV}$}
    \State Load joint model for $(S, GHI)$ calibrated to solar data
    \State Define production function $Q = Q(T, GHI)$ and set $P = GHI$
\EndIf
\Statex
\State Load contract specification
\If{$\mathcal{C} = \text{fixed-price}$}
    \State Define payoff $\psi(S) = S - K$
\ElsIf{$\mathcal{C} = \text{stepped}$}
    \State Define payoff $\psi(S) = S - K_l$, $l=1,...,L$
\ElsIf{$\mathcal{C} = \text{reverse collar}$}
    \State Define payoff $\psi(S) = (S - K_{max})^+ - (K_{min} - S)^+ + (S - K)\mathds{1}_{S\in (K_{min}, K_{max})}$
\EndIf
\Statex
\For{$path = 1$ to $N$}
    \State Simulate correlated shocks $\boldsymbol{\varepsilon}^{(path)} \sim \mathcal{N}(\mathbf{0}, \boldsymbol{\Sigma}_{\mathcal{T}})$
    \State Transform to $S^{(path)}$ and $P^{(path)}$  via calibrated models
    \State Compute $Q_{i,j}^{(path)} = Q(T_{i,j},P_{i,j}^{(path)})$
    \State Compute $\Pi_{path} = \sum_{i,j} Q_{i,j}^{(path)} \cdot \psi(S_{i,j}^{(path)})$
\EndFor
\State Sort $\{\Pi_{\text{path}}\}_{\text{path}=1}^N$ ascending: $\Pi_{(1)} \le \Pi_{(2)} \le \dots \le \Pi_{(N)}$
\State $k \gets \lfloor 0.05 N \rfloor$
\State $\widehat{\text{VaR}}_{0.95} \gets -\Pi_{(k)}$
\State $\widehat{\text{ES}}_{0.95} \gets -\frac{1}{k} \sum_{j=1}^{k} \Pi_{(j)}$
\end{algorithmic}
\end{algorithm}

\subsection{GHI modelling alternatives}\label{app:ghi_attempts}
In this Appendix we show the results of some alternative model specification attempts for GHI. For each model, we plot the original time series, the fitted seasonal values and the resulting non-seasonal components. What emerges from all figures is the inability to remove the strong seasonal pattern even by increasing the number of sine and cosine functions in the seasonal component.

\begin{figure}[H]
    \centering    \includegraphics[width=0.97\linewidth]{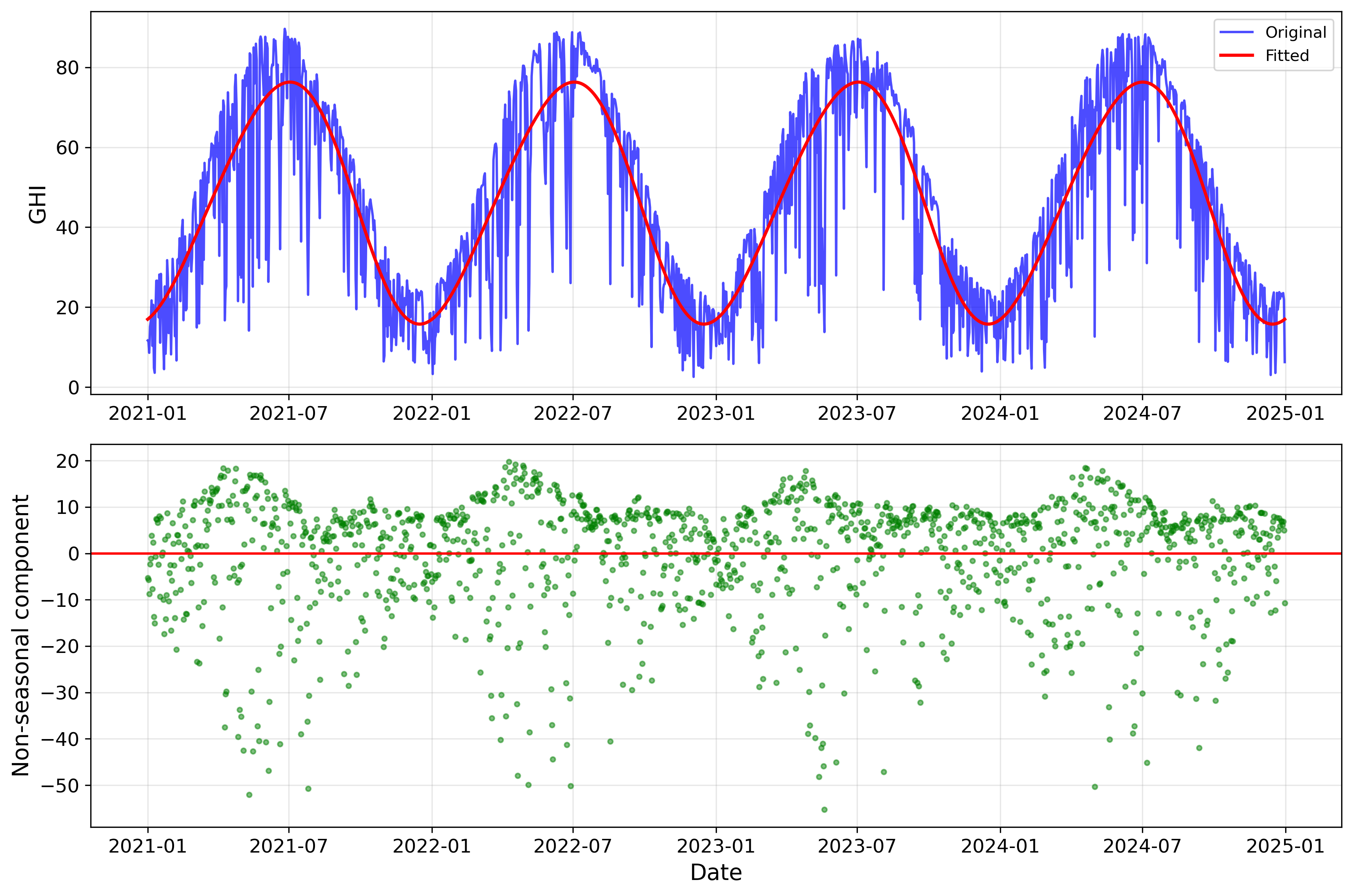}
    \caption{Additive model fit, $M=N=2$}
    \label{fig:additive_2}
\end{figure}

\begin{figure}[H]
    \centering    \includegraphics[width=0.97\linewidth]{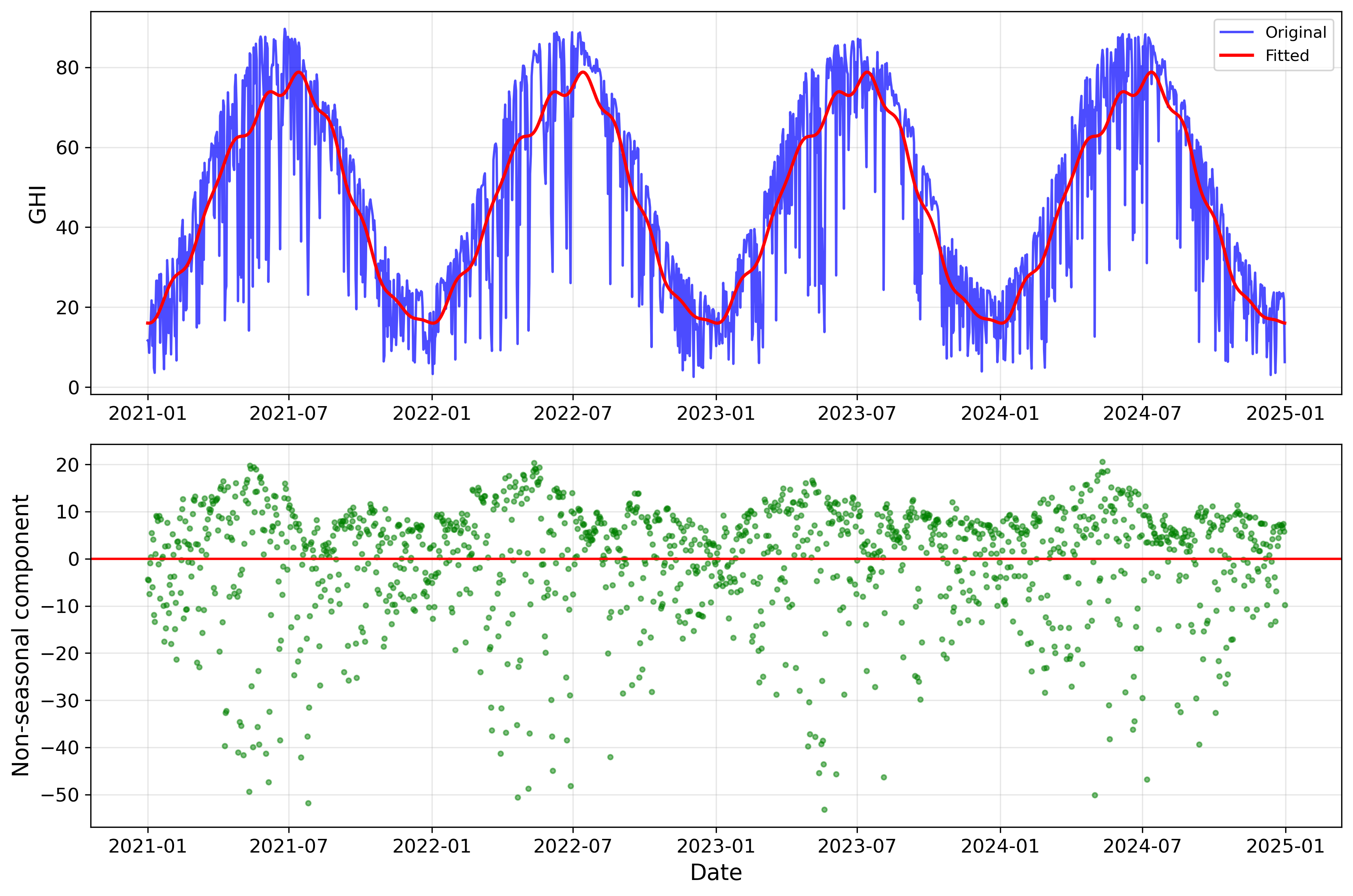}
    \caption{Additive model fit, $M=N=10$}
    \label{fig:additive_10}
\end{figure}

Model 1: additive, $M=N=2$ in Figure \ref{fig:additive_2} and $M=N=10$ in Figure \ref{fig:additive_10}.
\begin{align}
    \label{eq:solar_irradiance_additive}
    & GHI_t = \Lambda^{G}_t + Z_t
    \\
    \label{eq:seasonal_part_solar_additive}
    & \Lambda^{G}_t = a^G_0+ \sum_{k=1}^N a^G_k \sin\bigg(\frac{2\pi t}{365} \bigg) + \sum_{k=1}^M b^G_k \cos\bigg(\frac{2\pi t}{365} \bigg) 
    \\
    \label{eq:non_seasonal_part_solar_additive}
    & dZ_t = \gamma(\bar{Z} - Z_t) dt + \sigma_z dW_t^{G} + dL_t.
\end{align}

\begin{figure}[H]
    \centering    \includegraphics[width=0.97\linewidth]{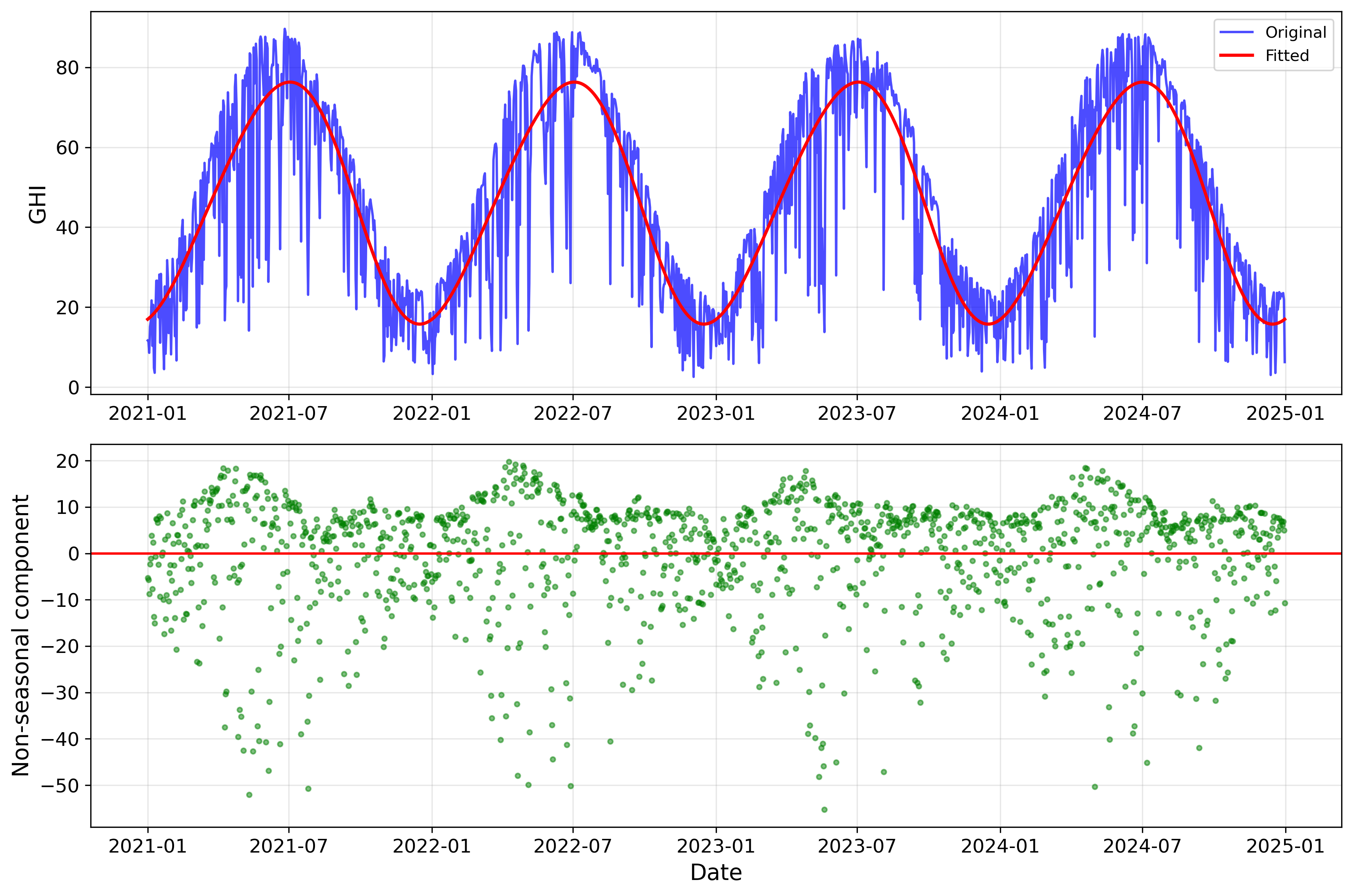}
    \caption{Multiplicative model fit, $M=N=2$}
    \label{fig:multiplicative_2}
\end{figure}

\begin{figure}[H]
    \centering    \includegraphics[width=0.97\linewidth]{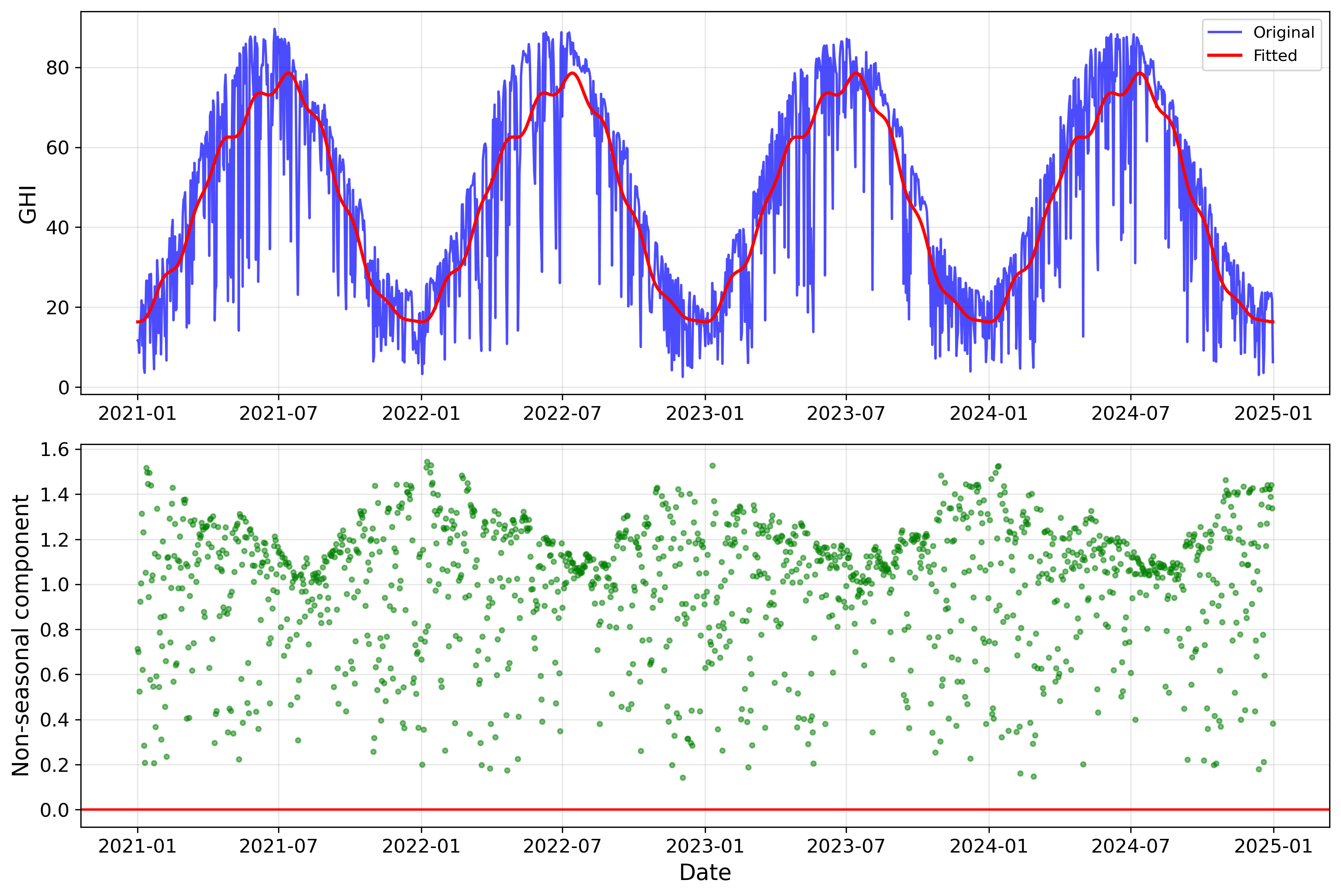}
    \caption{Multiplicative model fit, $M=N=10$}
    \label{fig:multiplicative_10}
\end{figure}

Model 2: multiplicative, $M=N=2$ in Figure \ref{fig:multiplicative_2} and $M=N=10$ in Figure \ref{fig:multiplicative_10}.
\begin{align}
    \label{eq:solar_irradiance_multiplicative}
    & GHI_t = \Lambda^{G}_t  Z_t
    \\
    \label{eq:seasonal_part_solar_multiplicative}
    & \Lambda^{G}_t = a^G_0+ \sum_{k=1}^N a^G_k \sin\bigg(\frac{2\pi t}{365} \bigg) + \sum_{k=1}^M b^G_k \cos\bigg(\frac{2\pi t}{365} \bigg) 
    \\
    \label{eq:non_seasonal_part_solar_multiplicative}
    & dZ_t = \gamma(\bar{Z} - Z_t) dt + \sigma_z dW_t^{G} + dL_t.
\end{align}

\begin{figure}[H]
    \centering    \includegraphics[width=0.97\linewidth]{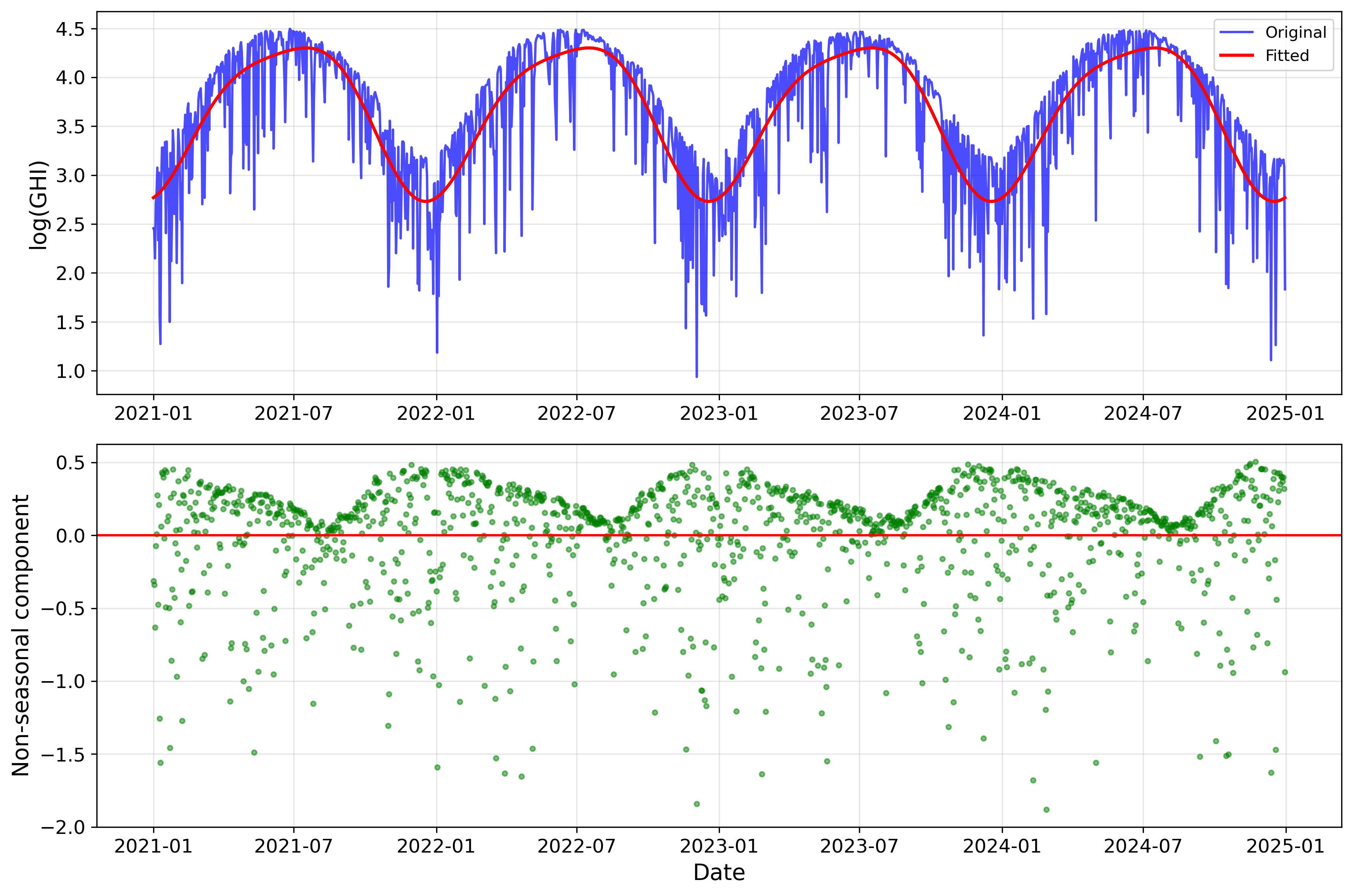}
    \caption{Exponential model fit, $M=N=2$}
    \label{fig:exponential_2}
\end{figure}

\begin{figure}[H]
    \centering    \includegraphics[width=0.97\linewidth]{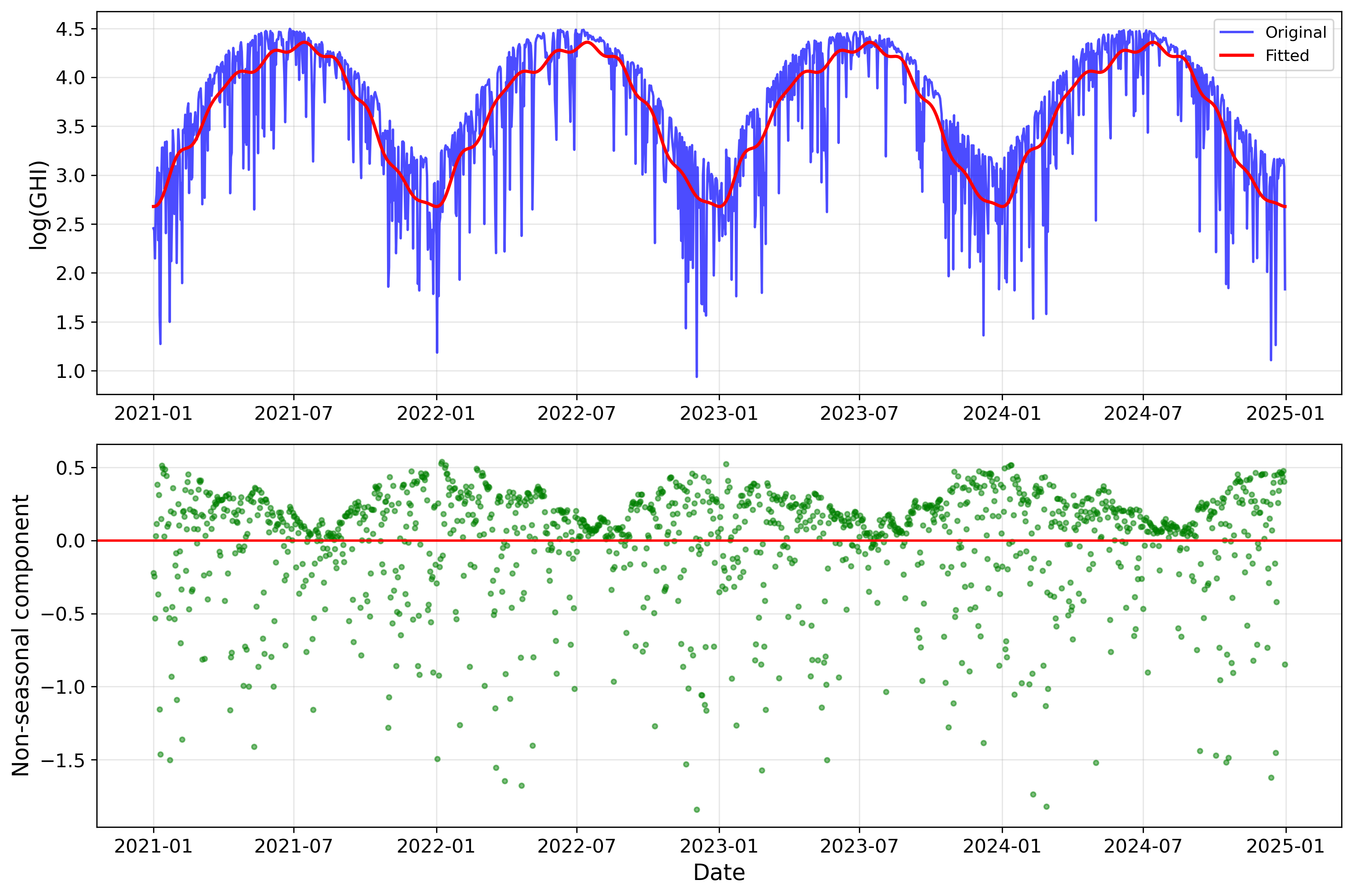}
    \caption{Exponential model fit, $M=N=10$}
    \label{fig:exponential_10}
\end{figure}

Model 3: exponential, $M=N=2$ in Figure \ref{fig:exponential_2} and $M=N=10$ in Figure \ref{fig:exponential_10}.
\begin{align}
    \label{eq:solar_irradiance_log}
    & GHI_t = \exp\{\Lambda^{G}_t + Z_t \}
    \\
    \label{eq:seasonal_part_solar_log}
    & \Lambda^{G}_t = a^G_0+ \sum_{k=1}^N a^G_k \sin\bigg(\frac{2\pi t}{365} \bigg) + \sum_{k=1}^M b^G_k \cos\bigg(\frac{2\pi t}{365} \bigg) 
    \\
    \label{eq:non_seasonal_part_solar_log}
    & dZ_t = \gamma(\bar{Z} - Z_t) dt + \sigma_z dW_t^{G} + dL_t.
\end{align}

\end{document}